\documentclass[aps,prd,onecolumn,superscriptaddress,nofootinbib,showpacs]{revtex4}

\usepackage{graphicx}
\usepackage{amssymb,amsmath,amsbsy}

\def\iso{\mathchoice{\cong}{\cong}{\isoS}{\cong}}
\def\isoS{\vbox{\baselineskip 0pt  \lineskip 0.5pt
    \ialign{$ \mathsurround=0pt  \scriptstyle \hfil ## \hfil $\crcr
        \sim \crcr = \crcr}}}

\begin{document}
\begin{flushright}
SCIPP-09/02\\[-1mm]
arXiv:0902.1537 [hep-ph]\\[-1mm]
\end{flushright}
\vspace*{1cm}

\title{Generalized CP symmetries and special regions of
parameter space in the two-Higgs-doublet model}

\author{P.\ M.\ Ferreira}
\affiliation{Instituto Superior de Engenharia de Lisboa,
    Rua Conselheiro Em\'{\i}dio Navarro,
    1900 Lisboa, Portugal}
\affiliation{Centro de F\'{\i}sica Te\'{o}rica e Computacional,
    Faculdade de Ci\^{e}ncias,
    Universidade de Lisboa,
    Av.\ Prof.\ Gama Pinto 2,
    1649-003 Lisboa, Portugal}
\author{Howard E.\ Haber}
\affiliation{Santa Cruz Institute for Particle Physics,
    University of California,
    Santa Cruz, California 95064, USA}
\author{Jo\~{a}o P.\ Silva}
\affiliation{Instituto Superior de Engenharia de Lisboa,
    Rua Conselheiro Em\'{\i}dio Navarro,
    1900 Lisboa, Portugal}
\affiliation{Centro de F\'{\i}sica Te\'{o}rica de Part\'{\i}culas,
    Instituto Superior T\'{e}cnico,
    P-1049-001 Lisboa, Portugal}

\date{\today}

\begin{abstract}
We consider the impact of imposing generalized
CP symmetries on the Higgs sector of the two-Higgs-doublet
model, and identify three classes of symmetries.
Two of these classes constrain the scalar potential parameters
to an exceptional region of parameter space which respects either
a $Z_2$ discrete flavor symmetry or a $U(1)$ symmetry.  
We exhibit a basis-invariant quantity that distinguishes between these
two possible symmetries.
We also show that the consequences of imposing these two
classes of CP symmetry can be achieved by
combining Higgs Family symmetries,
and that this is not possible for the usual CP symmetry.
We comment on the vacuum structure and
on renormalization in the presence of these symmetries.
Finally,
we demonstrate that the standard CP symmetry can be used to
build all the models we identify,
including those based on  Higgs Family symmetries.
\end{abstract}

\pacs{11.30.Er, 12.60.Fr, 14.80.Cp, 11.30.Ly}

\maketitle

\section{\label{sec:intro}Introduction}

Despite the fantastic successes of the Standard Model (SM)
of electroweak interactions,
its scalar sector remains largely untested~\cite{hhg}.
An alternative to the single Higgs doublet of the SM
is provided by the two-Higgs-doublet model (THDM),
which can be supplemented by symmetry requirements
on the Higgs fields $\Phi_1$ and $\Phi_2$.
Symmetries leaving the kinetic terms unchanged\footnote{It has
been argued by Ginsburg~\cite{Gin} and by
Ivanov~\cite{Ivanov1,Ivanov2} that one should also consider
the effect of non-unitary global symmetry transformations
of the two Higgs fields, as the most general renormalizable
Higgs Lagrangian allows for kinetic mixing of the two Higgs fields.
In this work, we study the possible
global symmetries of the effective low-energy Higgs theory that arise
\textit{after} diagonalization of the Higgs kinetic energy terms.
The non-unitary transformations that diagonalize the Higgs kinetic
mixing terms also transform the parameters of the Higgs potential,
and thus can determine the structure of the remnant Higgs flavor symmetries
of effective low-energy Higgs scalar potential.  It is the latter 
that constitutes the main focus of this work.}
may be of two types.  On the one hand,
one may relate $\Phi_a$ with some unitary transformation
of $\Phi_b$.
These are known as Higgs Family symmetries, or HF symmetries.
On the other hand,
one may relate $\Phi_a$ with some unitary transformation
of $\Phi_b^\ast$.
These are known as generalized CP symmetries,
or GCP symmetries.
In this article we consider all such symmetries that are
possible in the THDM,
according to their impact on the Higgs potential.
We identify three classes of GCP symmetries.

The study is complicated by the fact that one may perform a basis
transformation on the Higgs fields, thus hiding what might otherwise
be an easily identifiable symmetry. The need to seek basis invariant
observables in models with many Higgs was pointed out by Lavoura and
Silva \cite{LS}, and by Botella and Silva \cite{BS}, stressing
applications to CP violation. Refs.~\cite{BS,BLS} indicate how to
construct basis invariant quantities in a systematic fashion for any
model, including multi-Higgs-doublet models. Work on basis
invariance in the THDM was much expanded upon
by Davidson and Haber \cite{DavHab},
by Gunion and Haber \cite{GunHab,Gun},
by Haber and O'Neil \cite{HabONe},
and by other authors \cite{others}.
The previous approaches highlight the role
played by the Higgs fields. An alternative approach, spearheaded by
Nishi \cite{Nishi1,Nishi2}, by Ivanov \cite{Ivanov1,Ivanov2} and by
Maniatis {\em et al}~\cite{mani}, highlights the role played by
field bilinears, which is very useful for studies of the vacuum
structure of the model \cite{Barroso,earlier_bilinears}. In this
paper, we describe all classes of HF and GCP symmetries in both
languages.
One problem with two classes of GCP identified here is that
they lead to an exceptional region of parameter space
(ERPS) previously identified as problematic
by Gunion and Haber \cite{GunHab} and
by Davidson and Haber \cite{DavHab}.
Indeed, no basis invariant quantity exists in the literature that
distinguishes between the $Z_2$ and $U(1)$ HF symmetries in
the ERPS.

If evidence for THDM physics is revealed in future experiments, then
it will be critical to employ analysis techniques that are free
from model-dependent assumptions.  It is for this reason that
a basis-independent formalism for the THDM is so powerful.
Nevertheless, current experimental data already impose significant
constraints on the most general THDM.   In particular, we know that
custodial symmetry breaking effects, flavor changing neutral
current (FCNC) constraints, and (to a lesser extent) CP-violating phenomena 
impose some significant restrictions on the structure of the THDM
(including the Higgs-fermion interactions).  For example,
the observed suppression of FCNCs implies that either the two heaviest
neutral Higgs bosons of the THDM have masses above 1 TeV,
or certain Higgs-fermion Yukawa couplings must be absent~\cite{Pas}.
The latter can be achieved by imposing certain discrete symmetries on
the THDM.  Likewise, in the most general THDM, mass splittings between
charged and neutral Higgs bosons can yield custodial-symmetry breaking
effects at one-loop that could be large enough to be in 
conflict with the precision electroweak data~\cite{precision}.  
Once again, symmetries
can be imposed on the THDM to alleviate any potential disagreement
with data.  The implications of such symmetries for THDM phenomenology
has recently been explored by Gerard and collaborators~\cite{Gerard}
and by Haber and O'Neil~\cite{custodial}.

Thus, if THDM physics is discovered, it will be important to 
develop experimental methods that can reveal the presence or absence
of underlying symmetries of the most general THDM.  This requires two
essential pieces of input.  First, one must identify all possible
Higgs symmetries of interest.  Second, one must relate these
symmetries to basis-independent observables that can be probed by
experiment.  In this paper, we primarily address the first step,
although we also provide basis-independent characterizations of these
symmetries.  Our analysis focuses the symmetries of the THDM scalar
potential.  In principle, one can extend our study of these symmetries to the
Higgs-fermion Yukawa interactions, although this lies beyond the scope
of the present work.

This paper is organized as follows.
In section~\ref{sec:notation} we
introduce our notation and define an invariant that does
distinguish the $Z_2$ and $U(1)$ HF symmetries in
the ERPS.
In section~\ref{sec:vacuum} we explain the role
played by the vacuum expectation values in
preserving or breaking the $U(1)$ symmetry,
and we comment briefly on renormalization.
In section~\ref{sec:GCP} we introduce the GCP
transformations and explain why they are organized
into three classes.
We summarize our results and set them in the
context of the existing literature in section~\ref{sec:summary},
and
in section~\ref{sec:allisCP} we prove a surprising result:
multiple applications of
the standard CP symmetry can be used to
build all the models we identify,
including those based on  HF symmetries.
We draw our conclusions in
section~\ref{sec:conclusions}.

\section{\label{sec:notation}The scalar sector of the THDM}

\subsection{Three common notations for the scalar potential}

Let us consider a $SU(2) \otimes U(1)$ gauge theory with
two Higgs-doublets $\Phi_a$,
with the same hypercharge $1/2$,
and with vacuum expectation values (vevs)
\begin{equation}
\langle \Phi_a \rangle
=
\left(
\begin{array}{c}
0\\
v_a/\sqrt{2}
\end{array}
\right).
\label{vev}
\end{equation}
The index $a$ runs from $1$ to $2$,
and we use the standard definition for the electric
charge,
whereby the upper components of the $SU(2)$ doublets are
charged and the lower components neutral.

The scalar potential may be written as
\begin{eqnarray}
V_H
&=&
m_{11}^2 \Phi_1^\dagger \Phi_1 + m_{22}^2 \Phi_2^\dagger \Phi_2
- \left[ m_{12}^2 \Phi_1^\dagger \Phi_2 + \textrm{H.c.} \right]
\nonumber\\[6pt]
&&
+ \tfrac{1}{2} \lambda_1 (\Phi_1^\dagger\Phi_1)^2
+ \tfrac{1}{2} \lambda_2 (\Phi_2^\dagger\Phi_2)^2
+ \lambda_3 (\Phi_1^\dagger\Phi_1) (\Phi_2^\dagger\Phi_2)
+ \lambda_4 (\Phi_1^\dagger\Phi_2) (\Phi_2^\dagger\Phi_1)
\nonumber\\[6pt]
&&
+ \left[
\tfrac{1}{2} \lambda_5 (\Phi_1^\dagger\Phi_2)^2
+ \lambda_6 (\Phi_1^\dagger\Phi_1) (\Phi_1^\dagger\Phi_2)
+ \lambda_7 (\Phi_2^\dagger\Phi_2) (\Phi_1^\dagger\Phi_2)
+ \textrm{H.c.}
\right],
\label{VH1}
\end{eqnarray}
where $m_{11}^2$, $m_{22}^2$, and $\lambda_1,\cdots,\lambda_4$
are real parameters.
In general,
$m_{12}^2$, $\lambda_5$, $\lambda_6$ and $\lambda_7$
are complex. ``H.c.''~stands for Hermitian conjugation.

An alternative notation,
useful for the construction of invariants
and championed by Botella and Silva \cite{BS} is
\begin{eqnarray}
V_H
&=&
Y_{ab} (\Phi_a^\dagger \Phi_b) +
\tfrac{1}{2}
Z_{ab,cd} (\Phi_a^\dagger \Phi_b) (\Phi_c^\dagger \Phi_d),
\label{VH2}
\end{eqnarray}
where Hermiticity implies
\begin{eqnarray}
Y_{ab} &=& Y_{ba}^\ast,
\nonumber\\
Z_{ab,cd} \equiv Z_{cd,ab} &=& Z_{ba,dc}^\ast.
\label{hermiticity_coefficients}
\end{eqnarray}
The extremum conditions are
\begin{equation}
\left[ Y_{ab}
+ Z_{ab,cd}\, v_d^\ast v_c \right]\ v_b = 0
\hspace{3cm}(\textrm{for\ } a = 1,2).
\label{stationarity_conditions}
\end{equation}
Multiplying by $v_a^\ast$ leads to
\begin{equation}
Y_{ab} (v_a^\ast v_b) = - Z_{ab,cd}\,
(v_a^\ast v_b)\, (v_d^\ast v_c).
\label{aux_1}
\end{equation}

One should be very careful when comparing Eqs.~(\ref{VH1})
and (\ref{VH2}) among different authors,
since the same symbol may be used for quantities
which differ by signs, factors of two, or complex conjugation.
Here we follow the definitions of Davidson and Haber
\cite{DavHab}.
With these definitions:
\begin{eqnarray}
Y_{11}=m_{11}^2, &&
Y_{12}=-m_{12}^2,
\nonumber \\
Y_{21}=-(m_{12}^2)^\ast && Y_{22}=m_{22}^2,
\label{ynum}
\end{eqnarray}
and
\begin{eqnarray}
Z_{11,11}=\lambda_1, && Z_{22,22}=\lambda_2,
\nonumber\\
Z_{11,22}=Z_{22,11}=\lambda_3, && Z_{12,21}=Z_{21,12}=\lambda_4,
\nonumber \\
Z_{12,12}=\lambda_5, && Z_{21,21}=\lambda_5^\ast,
\nonumber\\
Z_{11,12}=Z_{12,11}=\lambda_6, && Z_{11,21}=Z_{21,11}=\lambda_6^\ast,
\nonumber \\
Z_{22,12}=Z_{12,22}=\lambda_7, && Z_{22,21}=Z_{21,22}=\lambda_7^\ast.
\label{znum}
\end{eqnarray}

The previous two notations look at the Higgs fields $\Phi_a$ individually.
A third notation is used by Nishi \cite{Nishi1,Nishi2} and
Ivanov \cite{Ivanov1,Ivanov2},
who emphasize
the presence of field bilinears $(\Phi_a^\dagger \Phi_b)$
\cite{earlier_bilinears}.
Following Nishi \cite{Nishi1} we write:
\begin{equation}
V_H = M_\mu r_\mu + \Lambda_{\mu \nu} r_\mu r_\nu,
\label{VH3}
\end{equation}
where $\mu = 0,1,2,3$ and
\begin{eqnarray}
r_0 &=&
\frac{1}{2}
\left[(\Phi_1^\dagger \Phi_1) + (\Phi_2^\dagger \Phi_2) \right],
\nonumber\\
r_1 &=&
\frac{1}{2}
\left[(\Phi_1^\dagger \Phi_2) + (\Phi_2^\dagger \Phi_1) \right]
= \textrm{Re}\, (\Phi_1^\dagger \Phi_2),
\nonumber\\
r_2 &=&
- \frac{i}{2}
\left[(\Phi_1^\dagger \Phi_2) - (\Phi_2^\dagger \Phi_1) \right]
= \textrm{Im}\, (\Phi_1^\dagger \Phi_2),
\nonumber\\
r_3 &=&
\frac{1}{2}
\left[(\Phi_1^\dagger \Phi_1) - (\Phi_2^\dagger \Phi_2) \right].
\label{r_Ivanov}
\end{eqnarray}
In Eq.~(\ref{VH3}),
summation of repeated indices is
adopted with Euclidean metric.
This differs from Ivanov's notation \cite{Ivanov1,Ivanov2},
who pointed out that $r_\mu$ parametrizes the gauge orbits
of the Higgs fields,
in a space equipped with a Minkowski metric.

In terms of the parameters of Eq.~(\ref{VH1}),
the $4$-vector $M_\mu$ and $4 \times 4$ matrix $\Lambda_{\mu\nu}$ are written
respectively as:
\begin{equation}
M_\mu =
\left(
\begin{array}{cccc}
m_{11}^2 + m_{22}^2,
&
-2\, \textrm{Re}\, m_{12}^2,
&
2\, \textrm{Im}\, m_{12}^2,
&
m_{11}^2 - m_{22}^2
\end{array}
\right),
\label{M_mu}
\end{equation}
and
\begin{equation}
\Lambda_{\mu \nu} =
\left(
\begin{array}{cccc}
(\lambda_1+\lambda_2)/2 + \lambda_3\
&\,\,\,
\textrm{Re}\, (\lambda_6 + \lambda_7)
&\,\,\,
- \textrm{Im}\, (\lambda_6 + \lambda_7)
&\,\,\,
(\lambda_1 - \lambda_2)/2
\\
 \phantom{-}\textrm{Re}\, (\lambda_6 + \lambda_7)\
&\,\,\,
\lambda_4 + \textrm{Re}\, \lambda_5
&\,\,\,
- \textrm{Im}\, \lambda_5
&\,\,\,
 \phantom{-}\textrm{Re}\, (\lambda_6 - \lambda_7)\
\\
- \textrm{Im}\, (\lambda_6 + \lambda_7)\
&\,\,\,
- \textrm{Im}\, \lambda_5
&\,\,\,
\lambda_4 - \textrm{Re}\, \lambda_5
&\,\,\,
- \textrm{Im}\, (\lambda_6 - \lambda_7)\
\\
(\lambda_1 - \lambda_2)/2
&\,\,\,
\textrm{Re}\, (\lambda_6 - \lambda_7)
&\,\,\,
- \textrm{Im}\, (\lambda_6 - \lambda_7)
&\,\,\,
\ (\lambda_1+\lambda_2)/2 - \lambda_3
\end{array}
\right).
\label{Lambda_munu}
\end{equation}
Eq.~(\ref{VH3}) is related to Eq.~(\ref{VH2}) through
\begin{eqnarray}
M^\mu &=&
\sigma^\mu_{ab}\, Y_{ba},
\label{M_vs_Y}
\\
\Lambda^{\mu \nu} &=&
\tfrac{1}{2} Z_{ab,cd}\, \sigma^\mu_{ba} \sigma^\nu_{dc},
\label{Lambda_vs_Z}
\end{eqnarray}
where the matrices $\sigma^i$ are the three Pauli matrices,
and $\sigma^0$ is the $2 \times 2$ identity matrix.

\subsection{Basis transformations}

We may rewrite the potential in terms of new fields $\Phi^\prime_a$,
obtained from the original ones by a simple 
(global) basis transformation
\begin{equation}
\Phi_a \rightarrow \Phi_a^\prime = U_{ab} \Phi_b,
\label{basis-transf}
\end{equation}
where $U\in U(2)$ is a $2 \times 2$ unitary matrix.
Under this unitary basis transformation,
the gauge-kinetic terms are unchanged,
but the coefficients $Y_{ab}$ and $Z_{ab,cd}$ are transformed as
\begin{eqnarray}
Y_{ab} & \rightarrow &
Y^\prime_{ab} =
U_{a \alpha}\, Y_{\alpha \beta}\, U_{b \beta}^\ast ,
\label{Y-transf}
\\
Z_{ab,cd} & \rightarrow &
Z^\prime_{ab,cd} =
U_{a\alpha}\, U_{c \gamma}\,
Z_{\alpha \beta,\gamma \delta}\, U_{b \beta}^\ast \, U_{d \delta}^\ast ,
\label{Z-transf}
\end{eqnarray}
and the vevs are transformed as
\begin{equation}
v_a \rightarrow v_a^\prime = U_{a b} v_b.
\label{vev-transf}
\end{equation}
Thus,
the basis transformations $U$ may be utilized in order to absorb
some of the degrees of freedom of $Y$ and/or $Z$,
which implies that not all parameters of Eq.~(\ref{VH2})
have physical significance.

\subsection{\label{subsec:HFsymmetry}Higgs Family symmetries}

Let us assume that the scalar potential in
Eq.~(\ref{VH2}) has some explicit internal symmetry.
That is,
we assume that the coefficients of $V_H$ stay
\textit{exactly the same} under a transformation
\begin{equation}
\Phi_a \rightarrow \Phi_a^S = S_{ab} \Phi_b.
\label{S-transf-symmetry}
\end{equation}
$S$ is a unitary matrix,
so that the gauge-kinetic couplings
are also left invariant by this Higgs Family symmetry
(HF symmetry).
As a result of this symmetry,
\begin{eqnarray}
Y_{a b} & = &
Y^S_{a b} =
S_{a \alpha}\, Y_{\alpha \beta}\, S_{b \beta}^\ast ,
\label{Y-S}
\\
Z_{ab,cd} & = &
Z^S_{ab,cd} =
S_{a \alpha}\, S_{c \gamma}\,
Z_{\alpha \beta, \gamma \delta}\, S_{b \beta}^\ast \, S_{d \delta}^\ast .
\label{Z-S}
\end{eqnarray}
Notice that this is \textit{not} the situation considered
in Eqs.~(\ref{basis-transf})--(\ref{Z-transf}).
There,
the coefficients of the Lagrangian
\textit{do change}
(although the quantities that are physically
measurable are invariant with respect to any change of basis).
In contrast, Eqs.~(\ref{S-transf-symmetry})--(\ref{Z-S})
imply the existence of a HF symmetry $S$ of the scalar potential
that leaves the coefficients of $V_H$ unchanged.

The Higgs Family symmetry group must be a subgroup of full $U(2)$
transformation group of $2\times 2$ unitary 
matrices employed in Eq.~(\ref{basis-transf}).  Given the most
general THDM scalar potential, there is always a $U(1)$ subgroup
of $U(2)$ under which the scalar potential is invariant.
This is the global hypercharge $U(1)_Y$ symmetry group:
\begin{equation}
U(1)_Y:
\hspace{4ex}
\Phi_1 \rightarrow e^{i \theta} \Phi_1,
\hspace{4ex}
\Phi_2 \rightarrow e^{i \theta} \Phi_2,
\label{U1Y}
\end{equation}
where $\theta$ is an arbitrary angle (mod $2\pi$).  The invariance
under the global $U(1)_Y$ is trivially guaranteed by the 
invariance under the $SU(2)\otimes U(1)$ electroweak gauge symmetry.
\textit{Since the  global hypercharge $U(1)_Y$ is always present, we shall 
henceforth define the 
HF symmetries as those Higgs Family symmetries that are 
orthogonal to $U(1)_Y$.}

We now turn to the interplay between
HF symmetries and basis transformations.
Let us imagine that,
when written in the basis of fields $\Phi_a$,
$V_H$ has a symmetry $S$.
We then perform a basis transformation from
the basis $\Phi_a$ to the basis $\Phi^\prime_a$,
as given by Eq.~(\ref{basis-transf}).
Clearly,
when written in the new basis,
$V_H$ does \textit{not} remain invariant under $S$.
Rather, it will be invariant under
\begin{equation}
S^\prime = U S U^\dagger .
\label{S-prime}
\end{equation}
As we change basis,
the form of the potential changes
in a way that may obscure the presence
of a HF symmetry.   In particular, two HF symmetries
that naively look distinct
will actually yield precisely the same physical predictions
if a unitary matrix $U$ exists such that Eq.~\eqref{S-prime} is satisfied.

HF symmetries in the two-Higgs-doublet model (THDM) have a long
history.
In papers by Glashow and Weinberg and by Paschos~\cite{Pas}, the discrete
$Z_2$ symmetry was introduced,
\begin{equation}
Z_2:
\hspace{4ex}
\Phi_1 \rightarrow \Phi_1,
\hspace{4ex}
\Phi_2 \rightarrow - \Phi_2,
\label{Z2}
\end{equation}
in order to preclude flavour-changing neutral currents \cite{Pas}.
This is just the interchange
\begin{equation}
\Pi_2:
\hspace{4ex}
\Phi_1 \leftrightarrow \Phi_2,
\label{Pi2}
\end{equation}
seen in a different basis,
as shown by applying Eq.~(\ref{S-prime}) in the form
\begin{equation}
\left(
\begin{array}{cc}
0 & 1 \\
1 & 0 \\
\end{array}
\right)
=
\frac{1}{\sqrt{2}}
\left(
\begin{array}{cc}
1 & 1 \\
1 & -1 \\
\end{array}
\right)
\
\left(
\begin{array}{cc}
1 & 0 \\
0 & -1 \\
\end{array}
\right)
\
\frac{1}{\sqrt{2}}
\left(
\begin{array}{cc}
1 & 1 \\
1 & -1 \\
\end{array}
\right).
\label{Z2ToPi2}
\end{equation}
Peccei and Quinn~\cite{PQ} introduced the continuous $U(1)$ symmetry
\begin{equation}
U(1):
\hspace{4ex}
\Phi_1 \rightarrow e^{-i \theta} \Phi_1,
\hspace{4ex}
\Phi_2 \rightarrow e^{i \theta} \Phi_2,
\label{U1}
\end{equation}
true for any value of $\theta$,
in connection with the strong CP problem.
Of course,
a potential invariant under $U(1)$ is also invariant
under $Z_2$.

Finally, we examine the largest possible Higgs Family symmetry group
of the THDM, namely $U(2)$.  In this case, a basis transformation
would have no effect on the Higgs potential parameters.  Since
$\delta_{ab}$ is the only $U(2)$-invariant tensor, it follows that
\begin{eqnarray}
Y_{ab}&=& c_1 \delta_{ab}\,,\label{u2y}\\
Z_{ab,cd} &=& c_2 \delta_{ab} \delta_{cd} + c_3 \delta_{ad} \delta_{bc}\,,
\label{u2z}
\end{eqnarray}
where $c_1$, $c_2$ and $c_3$ are arbitrary real 
numbers.~\footnote{Note that there is no $\delta_{ac} \delta_{bd}$ term
contributing to $Z_{ab,cd}$, as such a term is not invariant under
the transformation of Eq.~(\ref{Z-transf}).}
One can easily check from Eqs.~(\ref{Y-transf}) and (\ref{Z-transf})
that the unitarity of $U$ implies that $Y'=Y$ and $Z'=Z$ 
for any choice of basis, as required
by the $U(2)$-invariance of the scalar potential.
Eqs.~(\ref{u2y}) and (\ref{u2z}) impose the following constraints on
the parameters of the THDM scalar potential (independently of the
choice of basis):
\begin{eqnarray}
m_{22}^2 = m_{11}^2,
& \hspace{4ex} &
m_{12}^2=0,
\nonumber\\
\lambda_1 = \lambda_2=\lambda_3+\lambda_4\,,
& \hspace{4ex} &
\lambda_5 = \lambda_6 = \lambda_7= 0\,.
\label{u2pot}
\end{eqnarray}
As there are no non-zero potentially complex scalar potential parameters,
the $U(2)$-invariant THDM is clearly CP-invariant.

As previously noted, the
$U(2)$ symmetry contains the global hypercharge
$U(1)_Y$ as a subgroup.  Thus, in order to identify
the corresponding HF symmetry that is orthogonal to $U(1)_Y$,
we first observe that
\begin{equation}
U(2)\iso SU(2)\otimes U(1)_Y/Z_2 \iso SO(3)\otimes U(1)_Y\,.
\end{equation}
To prove the above isomorphism, simply note that any $U(2)$ matrix can
be written as $U=e^{i\theta}\hat{U}$, where $\hat U\in SU(2)$.
To cover the full $U(1)_Y$ group, we must take $0\leq\theta<2\pi$.
But since both $\hat U$ and $-\hat U$ are elements of $SU(2)$
whereas $+1$ and $-1=e^{i\pi}$ are elements of $U(1)_Y$, we must
identify $\hat U$ and $-\hat U$ as the same group element in order not to
double cover the full $U(2)$ group.  The identification of $\hat U$
with $-\hat U$ in $SU(2)$ is isomorphic to $SO(3)$, using the well known
isomorphism $SO(3)\iso SU(2)/Z_2$.  Consequently, we have identified
SO(3) as the HF symmetry that constrains the scalar potential
parameters as indicated in Eq.(\ref{u2pot}).

The impact of these symmetries on the potential parameters
in Eq.~(\ref{VH1}) is shown in section~\ref{sec:summary}.
As mentioned above,
if one makes a basis change, the potential parameters
change and so does the explicit form of the symmetry and
of its implications.
For example,
Eq.~(\ref{Z2ToPi2}) shows that the symmetries $Z_2$ and $\Pi_2$
are related by a basis change.
However,
they have a different impact on the parameters in their
respective basis.
This can be seen explicitly in Table~\ref{master1}
of section~\ref{sec:summary}. One can also easily prove that
the existence of either the $Z_2$, $\Pi_2$ or Peccei-Quinn $U(1)$
symmetry is sufficient to guarantee the existence of a basis choice in
which all scalar potential parameters are real.  That is, 
the corresponding scalar Higgs sectors are explicitly CP-conserving.

Basis invariant signs of HF symmetries were discussed
extensively in Ref.~\cite{DavHab}.
Recently,
Ferreira and Silva~\cite{FS2} extended these methods to include
Higgs models with more than two Higgs doublets.

Consider first the THDM scalar potentials that are invariant under 
the so-called \textit{simple} HF symmetries of Ref.~\cite{FS2}.
We define a simple HF symmetry to be a symmetry
group $G$ with the following property: the requirement that the
THDM scalar potential is invariant under a particular element 
$g\in G$ (where $g\neq e$ and $e$ is the identity element)
is sufficient to guarantee invariance under the entire
group $G$.  The discrete cyclic group 
$Z_n=\{e\,,\,g\,,\,g^2\,,\,\ldots\,,\,g^{n-1}\}$,
where $g^n = e$,
is an example of a possible simple HF symmetry group.  
If we restrict the TDHM scalar potential to include terms of
dimension-four or less (e.g., the tree-level scalar potential of the THDM),
then one can show that the Peccei-Quinn $U(1)$ symmetry is also
a simple HF symmetry.  For example, consider the matrix
\begin{equation}
S=\left( \begin{array}{cc}
e^{- 2 i \pi/3} & 0\\
0 & e^{2 i \pi/3}
\end{array} \right)\,.
\end{equation}
Note that $S$ is an element of the cyclic sub-group
$Z_3=\{S\,,\,S^2\,,\,S^3=1\}$ of the
Peccei-Quinn $U(1)$ group.
As shown in Ref.~\cite{FS2}, the 
invariance of the tree-level THDM scalar potential
under $\Phi_a\to S_{ab}\Phi_b$ automatically implies the
invariance of the scalar potential under the full Peccei-Quinn $U(1)$
group.  In contrast, the maximal HF symmetry, $SO(3)$, introduced
above is not a simple HF symmetry, as there is no single element of
$S\in SO(3)$ such that invariance under  $\Phi_a\to S_{ab}\Phi_b$ 
guarantees invariance of the tree-level THDM 
scalar potential under the
full SO(3) group of transformations.

Typically, the simple HF
symmetries take on a simple form for a particular choice of basis
for the Higgs fields.
We summarize here a few of the results of Ref.~\cite{FS2}:
\begin{enumerate}
\item In the THDM, there are only two
\textit{independent} classes of \textit{simple} symmetries:
a discrete $Z_2$ flavor symmetry, and a continuous Peccei-Quinn $U(1)$
flavor symmetry.
\item Other discrete flavor symmetry groups $G$ that are subgroups of $U(1)$
are not considered independent.  That is, if $S\in G$ (where
$S\neq e$), then invariance under the
the discrete symmetry $\Phi\to S\Phi$
makes the scalar potential automatically invariant under the full
Peccei-Quinn $U(1)$ group;
\item In most regions of parameter space,
one can build quantities invariant under basis transformations
that detect these symmetries;
\item There exists a so-called exceptional region of parameter space (ERPS)
characterized by
\begin{eqnarray}
m_{22}^2 = m_{11}^2,
& \hspace{4ex} &
m_{12}^2=0,
\nonumber\\
\lambda_2 = \lambda_1,
& \hspace{4ex} &
\lambda_7 = - \lambda_6.
\label{ERPS}
\end{eqnarray}
As shown by Davidson and Haber \cite{DavHab},
a theory obeying these constraints does have a $Z_2$ symmetry,
but it may or not have a $U(1)$ symmetry.
Within the ERPS, the invariants in the literature cannot be
used to distinguish the two cases.
\end{enumerate}

The last statement above is a result of the following considerations.
In order to distinguish between $Z_2$ and $U(1)$,
Davidson and Haber \cite{DavHab} construct two
invariant quantities given by Eqs.~(46) and (50) of Ref.~\cite{DavHab}.
Outside the ERPS,
these quantities are zero if and only if $U(1)$ holds.
Unfortunately,
in the ERPS these quantities vanish automatically
independently of whether or not $U(1)$ holds.
Similarly,
Ferreira and Silva \cite{FS2} have
constructed invariants detecting HF symmetries.
But their use requires the existence of a matrix, obtained
by combining $Y_{ab}$ and $Z_{ab,cd}$,
that has two distinct eigenvalues.
This does not occur when the ERPS is due to a symmetry.
Finally, in the ERPS,
Ivanov \cite{Ivanov1} states that the symmetry might be
``$(Z_2)^2$ or $O(2)$''
[our $Z_2$ \textit{or} our $U(1)$]
and does not provide a way to distinguish the
two possible flavor symmetries \cite{oversight}.

Gunion and Haber \cite{GunHab} have shown that
the ERPS conditions of Eq.~(\ref{ERPS})
are basis independent;
if they hold in one basis, then they hold in any basis.
Moreover, for a model in the ERPS,
a basis may be chosen such that all parameters are 
real.\footnote{Given a scalar potential whose parameters satisfy
the ERPS conditions
with ${\rm Im}(\lambda_5^* \lambda_6^2)\neq 0$, the unitary matrix
required to transform into a basis in which all the scalar potential
parameters are real can be determined only by numerical means.}
Having achieved such a basis,
Davidson and Haber \cite{DavHab} demonstrate that
one may make one additional basis transformation
such that
\begin{eqnarray}
m_{22}^2 = m_{11}^2,
& \hspace{4ex} &
m_{12}^2=0,
\nonumber\\
\lambda_2 = \lambda_1,
& \hspace{4ex} &
\lambda_7 = \lambda_6 = 0,
\hspace{4ex} \textrm{Im}\, \lambda_5 = 0.
\label{ERPS2}
\end{eqnarray}
These conditions express the ERPS for a specific basis choice.

One might think that this is such a special region of parameter
space that it lacks any relevance.
However,
the fact that the conditions in Eq.~(\ref{ERPS}) hold in
\textit{any} basis is a good indication that a
symmetry may lie behind this condition.
Indeed,
as pointed out by Davidson and Haber \cite{DavHab},
combining the two symmetries $Z_2$ and $\Pi_2$
\textit{in the same basis} one is lead immediately to
the ERPS in the basis of Eq.~(\ref{ERPS2}).
Up to now,
we considered the impact of imposing
on the Higgs potential only one symmetry.
This was dubbed a simple symmetry.
Now we are considering the possibility that the
potential must remain invariant under one symmetry
and \textit{also} under a second symmetry;
this implies further constraints on the parameters
of the Higgs potential.
We refer to this possibility as a multiple symmetry.
As seen from Table~\ref{master1} of section~\ref{sec:summary},
imposing $Z_2$ and $\Pi_2$ in the same basis leads
to the conditions in Eq.~(\ref{ERPS2}).
Incidentally,
this example shows that a model which lies in the
ERPS,
is automatically invariant under $Z_2$.

In section~{\ref{sec:GCP}} we will show that
all classes of non-trivial CP transformations lead
directly to the ERPS,
reinforcing the importance of this particular region of
parameter space.

\subsection{Requirements for $U(1)$ invariance}

In the basis in which the $U(1)$ symmetry takes the form
of Eq.~(\ref{U1}),
the coefficients of the potential must obey
\begin{equation}
m^{\prime\,2}_{12}= 0,
\hspace{4ex}
\lambda^\prime_5 = \lambda^\prime_6 = \lambda^\prime_7 = 0.
\label{U1_conditions}
\end{equation}
Imagine that we have a potential of Eq.~(\ref{VH1})
in the ERPS:
$m_{11}^2 = m_{22}^2$,
$m_{12}^2=0$,
$\lambda_2=\lambda_1$,
and $\lambda_7 = - \lambda_6$.
We now wish to know whether a transformation $U$ may be chosen
such that the potential coefficients in the new basis
satisfy the $U(1)$ conditions in Eq.~(\ref{U1_conditions}).
Using the transformation rules in Eqs.~(A13)-(A23) of Davidson and
Haber \cite{DavHab},
we find that such a choice of $U$ is possible if and only if the
coefficients in the original basis satisfy
\begin{equation}
2 \lambda_6^3 - \lambda_5 \lambda_6(\lambda_1 - \lambda_3 - \lambda_4)
- \lambda_5^2 \lambda_6^\ast = 0,
\label{can_change_to_usual_U1}
\end{equation}
subject to the condition that $\lambda_5^\ast \lambda_6^2$ is real.

\subsection{\label{subsec:D}The D invariant}

Having established the importance of the ERPS
(as it can arise from a symmetry),
we will now build a basis invariant quantity that
can be used to detect the presence of a U(1)
symmetry in this special case.

The quadratic terms of the Higgs potential are always
insensitive to the difference between $Z_2$ and $U(1)$.
Moreover,
the matrix $Y$ is proportional to the unit matrix in the ERPS.
One must thus look at the quartic terms.
We were inspired by the expression of $\Lambda_{\mu \nu}$
in Eq.~(\ref{Lambda_munu}),
which appears in the works of Nishi \cite{Nishi1,Nishi2} and
Ivanov \cite{Ivanov1,Ivanov2}.
In the ERPS of Eq.~(\ref{ERPS}),
$\Lambda_{\mu \nu}$ breaks into a $1 \times 1$ block
($\Lambda_{00}$),
and a $3 \times 3$ block
($\tilde{\Lambda} = \left\{\Lambda_{ij}\right\}$; $i,j=1,2,3$).
A basis transformation $U$ belonging to $SU(2)$ on the $\Phi_a$ fields
corresponds to an orthogonal $SO(3)$ transformation
in the $r_i$ bilinears,
given by
\begin{equation}
O_{ij} = \hbox{$\frac{1}{2}$}\,\textrm{Tr} 
\left[ U^\dagger \sigma_i U \sigma_j \right].
\label{O}
\end{equation}
Any matrix $O$ of $SO(3)$ can be obtained by considering an
appropriate matrix $U$ of $SU(2)$
(unfortunately this property does not generalize for
models with more than two Higgs doublets).
A suitable choice of $O$ can be made that diagonalizes
the $3 \times 3$ matrix $\tilde{\Lambda}$,
thus explaining Eq.~(\ref{ERPS2}).
In this basis,
the difference between the usual choices for $U(1)$ and $Z_2$ corresponds
to the possibility that $\textrm{Re} \lambda_5$ might
vanish or not, respectively.

We will now show that,
once in the ERPS,
the condition for the existence of $U(1)$ is that
$\tilde{\Lambda}$ has two eigenvalues which are equal.
The eigenvalues of a $3 \times 3$ matrix are the
solutions to the secular equation
\begin{equation}
x^3 + a_2 x^2 + a_1 x + a_0 = 0,
\label{secular}
\end{equation}
where
\begin{eqnarray}
a_0 &=&
\det \tilde{\Lambda}
= - \tfrac{1}{3}  \textrm{Tr\,} (\tilde{\Lambda}^3)
- \tfrac{1}{6} ( \textrm{Tr\,} \tilde{\Lambda} )^3
+ \tfrac{1}{2} ( \textrm{Tr\,} \tilde{\Lambda} )
\textrm{Tr\,} (\tilde{\Lambda}^2)
\nonumber\\
&=&
- \tfrac{1}{3} Z_{ab,cd}
\left( Z_{dc,gh} Z_{hg,ba} - \tfrac{3}{2} Z^{(2)}_{dc} Z^{(2)}_{ba} \right)
+ \tfrac{1}{2} Z_{ab,cd} Z_{dc,ba}
\textrm{Tr\,} \left( Z^{(1)} - \tfrac{1}{2} Z^{(2)} \right)
\nonumber\\
& &
-\tfrac{1}{6} \left( \textrm{Tr\,} Z^{(1)}\right)^3
+\tfrac{1}{4} \left( \textrm{Tr\,} Z^{(1)}\right)^2 \textrm{Tr\,} Z^{(2)}
- \tfrac{1}{2} \textrm{Tr\,} Z^{(1)} \left( \textrm{Tr\,} Z^{(2)}\right)^2,
\\
a_1 &=&
\tfrac{1}{2}  ( \textrm{Tr\,} \tilde{\Lambda} )^2
- \tfrac{1}{2} \textrm{Tr\,} (\tilde{\Lambda}^2)
\nonumber\\
&=&
\tfrac{1}{2}
\left[
\left( \textrm{Tr\,} Z^{(1)}\right)^2
- \textrm{Tr\,} Z^{(1)} \textrm{Tr\,} Z^{(2)}
+ \left( \textrm{Tr\,} Z^{(2)}\right)^2
- Z_{ab,cd} Z_{dc,ba}
\right],
\\
a_2 &=&
- \textrm{Tr\,} \tilde{\Lambda}
\nonumber\\
&=&
\tfrac{1}{2} \textrm{Tr\,} Z^{(2)} - \textrm{Tr\,} Z^{(1)},
\end{eqnarray}
and
\begin{eqnarray}
Z_{ab}^{(1)}
&\equiv&
Z_{a \alpha,\alpha b} =
\left( \begin{array}{cc}
  \lambda_1 + \lambda_4 & \quad \lambda_6 + \lambda_7 \\
  \lambda_6^\ast + \lambda_7^\ast & \quad \lambda_2 + \lambda_4
\end{array} \right),
\\
Z_{ab}^{(2)}
&\equiv&
Z_{\alpha \alpha,a b} =
\left( \begin{array}{cc}
  \lambda_1 + \lambda_3 & \quad \lambda_6 + \lambda_7 \\
  \lambda_6^\ast + \lambda_7^\ast & \quad \lambda_2 + \lambda_3
\end{array} \right).
\end{eqnarray}
The cubic equation, Eq.~(\ref{secular}), has at least two
degenerate solutions if \cite{AbrSte}
\begin{equation}
D \equiv
\left[ \tfrac{1}{3} a_1 - \tfrac{1}{9} a_2^2 \right]^3
+  \left[ \tfrac{1}{6} (a_1 a_2 - 3 a_0) - \tfrac{1}{27} a_2^3 \right]^2
\end{equation}
vanishes.

The expression of $D$ in terms of the parameters in Eq.~(\ref{VH1})
is rather complicated,
even in the ERPS.
But one can show by direct computation that if the $U(1)$-symmetry
condition of Eq.~(\ref{can_change_to_usual_U1}) holds
(subject to $\lambda_5^\ast \lambda_6^2$ being real),
then $D=0$.
We can simplify the expression for $D$ by changing to a
basis where all parameters are real \cite{GunHab},
where we get
\begin{equation}
D =
- \tfrac{1}{27}
\left[ \lambda_5 (\lambda_1 - \lambda_3 - \lambda_4 + \lambda_ 5)
       - 2 \lambda_6^2 \right]^2
\left[ (\lambda_1 - \lambda_3 - \lambda_4 - \lambda_ 5)^2
       + 16 \lambda_6^2 \right].
       \label{eq:D}
\end{equation}
If $\lambda_6 \neq 0$, then $D=0$ means
\begin{equation}
2 \lambda_6^2 = \lambda_5 (\lambda_1 - \lambda_3 - \lambda_4 + \lambda_ 5).
\label{L6NEQ0}
\end{equation}
If $\lambda_6 = 0$,
then $D=0$ corresponds to one of three possible conditions:
\begin{equation}
\lambda_5 = 0,
\hspace{4ex}
\lambda_5 = \pm (\lambda_1 - \lambda_3 - \lambda_4).
\label{L6EQ0}
\end{equation}
Notice that Eqs.~(\ref{L6NEQ0}) and (\ref{L6EQ0}) are
equivalent to Eq.~(\ref{can_change_to_usual_U1})
in any basis where the coefficients are real.

Although $D$ can be defined outside the ERPS,
the condition $D=0$ only guarantees that the model is invariant under
$U(1)$ inside the ERPS of Eq.~(\ref{ERPS}).
Outside this region one can detect the presence of a $U(1)$ symmetry
with the invariants proposed by Davidson and Haber \cite{DavHab}.
This closes the last breach in the literature concerning basis-invariant
signals of discrete symmetries in the THDM.
Thus, in the ERPS $D=0$ is a necessary and sufficient condition for
the presence of a $U(1)$ symmetry.

\section{\label{sec:vacuum}Vacuum structure and renormalization}

The presence of a $U(1)$ symmetry in the Higgs potential
may (or not) imply the existence of a massless scalar, the axion,
depending on whether (or not) the $U(1)$ is broken by the vevs.
In the previous section we related the basis-invariant condition
$D=0$ in the ERPS with the presence of a $U(1)$ symmetry.
In this section we will show that,
whenever the basis-invariant condition $D=0$ is
satisfied in the ERPS,
there is always a stationary point for which a massless scalar,
other than the usual Goldstone bosons, exists.

We start by writing the extremum conditions for the THDM in the
ERPS.
For simplicity, we will be working in a basis where all the
parameters are real~\cite{GunHab}.
From Eqs.~\eqref{stationarity_conditions} and~\eqref{znum},
we obtain
\begin{align}
0 = & \;\;
Y_{11}\,v_1\,+\,\frac{1}{2}\,\left[\lambda_1\,v_1^3\,+
\,\lambda_{345}\,v_1\,v_2^2\,+\,
\lambda_6\,(3\,v_1^2\,v_2\,-\,v_2^3)\right],
\nonumber\\
0 = & \;\;
Y_{11}\,v_2\,+\,\frac{1}{2}\,\left[\lambda_1\,v_2^3\,+
\,\lambda_{345}\,v_2\,v_1^2\,+\,
\lambda_6\,(v_1^3\,-\,3\,v_2^2\,v_1)\right],
\label{eq:stat}
\end{align}
where we have defined
$\lambda_{345}\equiv\lambda_3 + \lambda_4 + \lambda_5$.
We now compute the mass matrices.
As we will be considering only vacua with real vevs,
there will be no mixing between the real and imaginary
parts of the doublets.
As such, we can define the mass matrix of the CP-even scalars as given by
\begin{equation}
\left[M^2_h\right]_{ij}\;=\;\frac{1}{2}\,\frac{\partial^2
V}{\partial \mbox{Re}(\Phi_i^0)\,
\partial \mbox{Re}(\Phi_j^0)}
\end{equation}
where $\Phi_i^0$ is the neutral (lower) component of the $\Phi_i$
doublet. Thus, we obtain, for the entries of this matrix, the
following expressions:
\begin{align}
\left[M^2_h\right]_{11}\;=& \;
Y_{11}\,+\,\frac{1}{2}\,\left(3\,\lambda_1\,v_1^2\,+
\,\lambda_{345}\,v_2^2
\,
+\,6\,\lambda_6\,v_1\,v_2\right)\nonumber \vspace{0.2cm} \\
\left[M^2_h\right]_{22}\;=&\;
Y_{11}\,+\,\frac{1}{2}\,\left(3\,\lambda_1\,v_2^2\,+
\,\lambda_{345}\,v_1^2
\,
+\,6\,\lambda_6\,v_1\,v_2\right)\nonumber \vspace{0.2cm} \\
\left[M^2_h\right]_{12}\;=& \;\lambda_{345}\,v_1\,v_2\,
\,+\,\frac{3}{2}\,\lambda_6\,(v_1^2\,-\,v_2^2)\;\;\; .
\label{eq:mh}
\end{align}
Likewise, the pseudoscalar mass matrix is defined as
\begin{equation}
\left[M^2_A\right]_{ij}\;=\;\frac{1}{2}\,\frac{\partial^2
V}{\partial \mbox{Im}(\Phi_i^0)
\partial \mbox{Im}(\Phi_j^0)}
\end{equation}
whose entries are given by
\begin{align}
\left[M^2_A\right]_{11}\;=&\;
Y_{11}\,+\,\frac{1}{2}\,\left[\lambda_1\,v_1^2\,+\,
\left(\lambda_3\,+\,\lambda_4\,-\,\lambda_5\right)\,v_2^2
\,+\,2\,\lambda_6\,v_1\,v_2 \right]
\nonumber \vspace{0.2cm} \\
\left[M^2_A\right]_{22}\;=&\;
Y_{11}\,+\,\frac{1}{2}\,\left[\lambda_1\,v_2^2\,+\,
\left(\lambda_3\,+\,\lambda_4\,-\,\lambda_5\right)\,v_1^2
\,-\,2\,\lambda_6\,v_1\,v_2 \right]
\nonumber \vspace{0.2cm} \\
\left[M^2_A\right]_{12}\;=& \;\lambda_5\,v_1\,v_2 \,+\,
\frac{1}{2}\,\lambda_6\,(v_1^2\,-\,v_2^2)\;\;\; .
\label{eq:mA}
\end{align}
The expressions ~\eqref{eq:mh} and~\eqref{eq:mA} are valid for all
the particular cases we will now consider.

\subsection{Case $\lambda_6\,=\,0$, $\{v_1\,,\,v_2\}\,\neq\,0$}

Let us first study the case $\lambda_6\,=\,0$, wherein we may solve
the extremum conditions in an analytical manner. It is trivial
to see that Eqs.~\eqref{eq:stat} have three types of solutions: both
vevs different from zero, one vev equal to zero (say, $v_2$) and
both vevs zero (trivial non-interesting solution). For a solution
with $\{v_1\,,\,v_2\}\,\neq\,0$, a necessary condition must be
obeyed so that there is a solution to Eqs.~\eqref{eq:stat}:
\begin{equation}
\lambda_1^2\,-\,\lambda_{345}^2 \;\neq\;0\;\;\; . \label{eq:det}
\end{equation}
If we use the extremum conditions to evaluate
$\left[M^2_h\right]$, we obtain
\begin{equation}
\left[M^2_h\right]\;=\;\begin{pmatrix} \lambda_1\,v_1^2 & \quad
\lambda_{345}\,v_1\,v_2 \\ \lambda_{345}\,v_1\,v_2 & \quad
\lambda_1\,v_2^2
\end{pmatrix}
\end{equation}
which only has a zero eigenvalue if Eq.~\eqref{eq:det} is broken.
Thus, there is no axion in this matrix in this case. As for
$\left[M^2_A\right]$, we get
\begin{equation}
\left[M^2_A\right]\;=\;-\,\lambda_5\,\begin{pmatrix}
v_1^2 &\quad  v_1\,v_2 \\
v_1\,v_2 &\quad v_2^2 \end{pmatrix}
\end{equation}
which clearly has a zero eigenvalue corresponding to the $Z$
Goldstone boson. Further, this matrix will have an axion if
$\lambda_5\,=\,0$, which is the first condition of
Eq.~\eqref{L6EQ0}.

\subsection{Case $\lambda_6\,=\,0$, $\{v_1\,\neq\,0,\,v_2\,=\,0\}\,$}

Returning to Eq.~\eqref{eq:stat}, this case gives us
\begin{equation}
Y_{11}\,=\,-\,\frac{1}{2}\,\lambda_1\,v_1^2\;\;\; ,
\end{equation}
which implies $Y_{11}\,<\,0$. With this condition, the mass matrices
become considerably simpler:
\begin{equation}
\left[M^2_h\right]\;=\;\begin{pmatrix} \lambda_1\,v_1^2 & \quad 0 \\
 0 & \quad \frac{1}{2}\,(\lambda_{345}\,-\,\lambda_1)\,v_1^2
\end{pmatrix}
\label{eq:mhll}
\end{equation}
and
\begin{equation}
\left[M^2_A\right]\;=\;\frac{1}{2}\,\begin{pmatrix} 0 & \quad 0 \\
 0 & \quad (\lambda_3\,+\,\lambda_4\,-\lambda_5\,-\,\lambda_1)\,v_1^2
\end{pmatrix} \;\;\; .
\label{eq:mAll}
\end{equation}
So, we can have an axion in the matrix~\eqref{eq:mhll} if
\begin{equation}
\lambda_{345}\,-\,\lambda_1\,=\,0\;\;\Leftrightarrow\;\;\lambda_5\,=\,
\lambda_1\,-\,\lambda_3\,-\,\lambda_4 \label{eq:c1}
\end{equation}
or an axion in matrix~\eqref{eq:mAll} if
\begin{equation}
\lambda_5\,=\, -\lambda_1\,+\,\lambda_3\,+\,\lambda_4 \;\;\; .
\label{eq:c2}
\end{equation}
That is, we have an axion if the second or third conditions of
Eq.~\eqref{L6EQ0} are satisfied. The other possible case,
$\{v_1\,=\,0,\,v_2\,\neq\,0\}\,$, produces exactly the same
conclusions.

\subsection{Case $\lambda_6\,\neq\,0$}

This is the hardest case to treat, since we cannot obtain analytical
expressions for the vevs. Nevertheless a full analytical treatment is
still possible. First, notice that with $\lambda_6\,\neq\,0$
Eqs.~\eqref{eq:stat} imply that both vevs have to be non-zero. At
the stationary point of Eqs.~\eqref{eq:stat}, the pseudoscalar mass
matrix has a Goldstone boson and an eigenvalue given by
\begin{equation}
-\lambda_5\,(v_1^2\,+\,v_2^2)\,-
\,\lambda_6\,\frac{v_1^4\,-v_2^4}{2\,v_1\,v_2}
\;\;\; .
\end{equation}
So, an axion exists if we have
\begin{equation}
\frac{v_1^2\,-\,v_2^2}{v_1\,v_2}\;=\;-
\,\frac{2\,\lambda_5}{\lambda_6}\;\;\;.
\label{eq:ves}
\end{equation}
On the other hand, after some algebraic manipulation, it is simple
to obtain from~\eqref{eq:stat} the following condition:
\begin{equation}
\lambda_1\,-\,\lambda_{345}\;=\;
\lambda_6\,\left(\frac{v_1^2\,-\,v_2^2}{v_1\,v_2}\,-
\,\frac{4\,v_1\,v_2}{v_1^2\,-\,v_2^2}\right)
\label{eq:les}
\end{equation}
Substituting Eq.~\eqref{eq:ves} into~\eqref{eq:les}, we obtain
\begin{equation}
\lambda_1\,-\,\lambda_{345}\,=\,
\lambda_6\,\left(-\,\frac{2\,\lambda_5}{\lambda_6}\,+
\,\frac{2\,\lambda_6}{\lambda_5}\right)
\;\Longleftrightarrow\; 2 \lambda_6^2 \,=
\, \lambda_5 (\lambda_1 -
\lambda_3 - \lambda_4 + \lambda_ 5).
\end{equation}

Thus, we have shown that all of the conditions stemming from the
basis-invariant condition $D=0$
guarantee the existence of some stationary point for
which the scalar potential yields an axion.
Notice that, however,
this stationary point need not coincide with the global minimum
of the potential.

\subsection{Renormalization group invariance}

We now briefly examine the renormalization group (RG) behavior of our
basis-invariant condition $D=0$. It would be meaningless to say that
$D=0$ implies a $U(1)$ symmetry if that condition were only valid at
a given renormalization scale.  That is, it could well be that a numerical
accident forces $D=0$ at only a given scale. To avoid such a
conclusion, we must verify if
$D=0$ is a RG-invariant condition (in addition to being
basis-invariant). For a given renormalization scale $\mu$, the
$\beta$-function of a given parameter $x$ is defined as
$\beta_x\,=\,\mu\,\partial x/\partial \mu$. For simplicity, let us
rewrite $D$ in Eq.~\eqref{eq:D} as
\begin{equation}
D\;=\;-\,\frac{1}{27}\,D_1^2\,D_2\;\;\; ,
\end{equation}
with
\begin{align}
D_1 &=\; \lambda_5 (\lambda_1 - \lambda_3 - \lambda_4 + \lambda_ 5)
       - 2 \lambda_6^2 \nonumber \\
D_2 &=\; (\lambda_1 - \lambda_3 - \lambda_4 - \lambda_ 5)^2
       + 16 \lambda_6^2 \;\;\;.
\end{align}
If we apply the operator $\mu\,\partial /\partial \mu$ to $D$, we
obtain
\begin{equation}
\beta_D\,=\,-\,\frac{1}{27}\,
\left(2\,D_1\,D_2\,\beta_{D_1}\,+\,D_1^2\,\beta_{D_2}\right)
\;\;\;.
\end{equation}

If $D_1=0$ (which corresponds to three of the conditions presented
in Eqs.~\eqref{L6NEQ0} and~\eqref{L6EQ0}) then we immediately have
$\beta_D\,=\,0$.  That is, if $D=0$ at a given scale, it is zero at
all scales.

If $D_2=0$ and $D_1\neq 0$ we will only have $\beta_D\,=\,0$ if
$\beta_{D_2}\,=\,0$, or equivalently,
\begin{equation}
2\,(\lambda_1 - \lambda_3 - \lambda_4 - \lambda_ 5)\,
(\beta_{\lambda_1} - \beta_{\lambda_3} - \beta_{\lambda_4} -
\beta_{\lambda_ 5}) \,+\,32\,\beta_{\lambda_6}\,\lambda_6\;=\;0
\;\;\; .
\end{equation}
Given that $D_2=0$ implies that $\lambda_6\,=\,0$ and
$\lambda_5\,=\,\lambda_1 - \lambda_3 - \lambda_4$, we once
again obtain $\beta_D\,=\,0$.

Thus, the condition $D=0$ is RG-invariant. A direct verification of
the RG invariance of Eqs.~\eqref{L6NEQ0} and~\eqref{L6EQ0}, and of
the conditions that define the ERPS itself, would require the
explicit form of the $\beta$ functions of the THDM involving the
$\lambda_6$ coupling. That verification will be made
elsewhere~\cite{drtj}.

\section{\label{sec:GCP}Generalized CP symmetries}

It is common to consider the standard CP transformation
of the scalar fields as
\begin{equation}
\Phi_a (t, \vec{x}) \rightarrow
\Phi^{\textrm{CP}}_a (t, \vec{x}) = \Phi_a^\ast (t, - \vec{x}),
\label{StandardCP}
\end{equation}
where the reference to the time ($t$) and space ($\vec{x}$)
coordinates will henceforth be suppressed.
However,
in the presence of several scalars with the same quantum numbers,
basis transformations can be included in the definition of the
CP transformation.
This yields generalized CP transformations (GCP),
\begin{eqnarray}
\Phi^{\textrm{GCP}}_a
&=& X_{a \alpha} \Phi_\alpha^\ast
\equiv X_{a \alpha} (\Phi_\alpha^\dagger)^\top,
\nonumber\\
\Phi^{\dagger \textrm{GCP}}_a
&=& X_{a \alpha}^\ast \Phi_\alpha^\top
\equiv X_{a \alpha}^\ast (\Phi_\alpha^\dagger)^\ast,
\label{GCP}
\end{eqnarray}
where $X$ is an arbitrary unitary
matrix~\cite{GCP1,GCP2}.\footnote{Equivalently, one can 
consider a generalized time-reversal
transformation proposed in Ref.~\cite{Branco:1983tn}
and considered further in Appendix A of Ref.~\cite{GunHab}.}

Note that the transformation
$\Phi_a\to\Phi^{\rm GCP}_a$, where $\Phi^{\rm GCP}_a$ is given
by Eq.~\eqref{GCP},
leaves the kinetic terms invariant.
The GCP transformation of a field bilinear yields
\begin{equation}
\Phi^{\dagger \textrm{GCP}}_a
\Phi^{\textrm{GCP}}_b
=
X_{a \alpha}^\ast X_{b \beta}
(\Phi_\alpha \Phi_\beta^\dagger)^\top.
\end{equation}
Under this GCP transformation,
the quadratic terms of the potential may be written as
\begin{eqnarray}
Y_{ab} \Phi^{\dagger \textrm{GCP}}_a
\Phi^{\textrm{GCP}}_b
&=&
Y_{ab} X_{a \alpha}^\ast X_{b \beta}
\Phi_\beta^\dagger \Phi_\alpha
\nonumber\\
&=&
X_{b \beta} Y_{ba}^\ast X_{a \alpha}^\ast
\Phi_\beta^\dagger \Phi_\alpha
\nonumber\\
&=&
X_{\alpha a} Y_{\alpha \beta}^\ast X_{\beta b}^\ast
\Phi_a^\dagger \Phi_b
=
( X^\dagger\, Y\, X )^\ast_{ab}
\Phi_a^\dagger \Phi_b.
\end{eqnarray}
We have used the Hermiticity condition
$Y_{ab}=Y_{ba}^\ast$ in going to the second line;
and changed the dummy indices $a \leftrightarrow \beta$
and $b \leftrightarrow \alpha$ in going to the third line.
A similar argument can be made for the quartic terms.
We conclude that the potential is invariant
under the GCP transformation
of Eq.~\eqref{GCP} if and only if the coefficients obey
\begin{eqnarray}
Y_{ab}^\ast
&=&
X_{\alpha a}^\ast Y_{\alpha \beta} X_{\beta b}
= ( X^\dagger\, Y\, X )_{ab},
\nonumber\\
Z_{ab,cd}^\ast
&=&
X_{\alpha a}^\ast X_{\gamma c}^\ast
Z_{\alpha \beta, \gamma \delta} X_{\beta b} X_{\delta d}.
\label{YZ-CPtransf}
\end{eqnarray}

Introducing
\begin{eqnarray}
\Delta Y_{ab}
&=&
Y_{ab} -
X_{\alpha a} Y_{\alpha \beta}^\ast X_{\beta b}^\ast
= \left[Y - ( X^\dagger\, Y\, X )^\ast \right]_{ab},
\nonumber\\
\Delta Z_{ab,cd}
&=&
Z_{ab,cd} -
X_{\alpha a} X_{\gamma c}
Z_{\alpha \beta, \gamma \delta}^\ast X_{\beta b}^\ast X_{\delta d}^\ast.
\label{DY-DZ}
\end{eqnarray}
we may write the conditions for invariance under GCP as
\begin{eqnarray}
\Delta Y_{ab}
&=&
0,
\label{DY-GCP}
\\
\Delta Z_{ab,cd}
&=&
0.
\label{DZ-GCP}
\end{eqnarray}
Given Eqs.~\eqref{hermiticity_coefficients},
it is easy to show that
\begin{eqnarray}
\Delta Y_{ab} &=& \Delta Y_{ba}^\ast,
\nonumber\\
\Delta Z_{ab,cd} \equiv \Delta Z_{cd,ab} &=& \Delta Z_{ba,dc}^\ast.
\label{DY-DZ-hermiticity}
\end{eqnarray}
Thus, we need only consider the real coefficients
$\Delta Y_{11}$, $\Delta Y_{22}$,
$\Delta Z_{11,11}$, $\Delta Z_{22,22}$,
$\Delta Z_{11,22}$, $\Delta Z_{12,21}$,
and the complex coefficients
$\Delta Y_{12}$, $\Delta Z_{11,12}$,
$\Delta Z_{22,12}$, and $\Delta Z_{12,12}$.

\subsection{GCP and basis transformations}

We now turn to the interplay between GCP transformations and basis
transformations.
Consider the potential of Eq.~\eqref{VH2} and call it
$V(\Phi)$.
Now consider the potential obtained from $V(\Phi)$
by the basis transformation
$\Phi_a \rightarrow \Phi^\prime_a = U_{ab} \Phi_b$:
\begin{equation}
V (\Phi^\prime) =
Y_{ab}^\prime (\Phi_a^{\prime \dagger} \Phi^\prime_b) +
\tfrac{1}{2}
Z^\prime_{ab,cd} (\Phi_a^{\prime \dagger} \Phi^\prime_b)
(\Phi_c^{\prime \dagger} \Phi^\prime_d),
\end{equation}
where the coefficients in the new basis are given by
Eqs.~(\ref{Y-transf}) and (\ref{Z-transf}).
We will now prove the following theorem: If $V(\Phi)$ is invariant under
the GCP transformation of Eq.~(\ref{GCP}) with the matrix $X$,
then $V (\Phi^\prime)$ is invariant under a new GCP transformation
with matrix
\begin{equation}
X^\prime = U X U^\top.
\label{X-prime}
\end{equation}
By hypothesis $V(\Phi)$ is invariant under
the GCP transformation of Eq.~(\ref{GCP}) with the matrix $X$.
Eq.~(\ref{YZ-CPtransf}) guarantees that $Y^\ast = X^\dagger Y X$.
Now,
Eq.~(\ref{Y-transf}) relates the coefficients in the two
basis through $Y = U^\dagger Y^\prime U$.
Substituting gives
\begin{equation}
U^\top Y^{\prime \ast} U^\ast
= X^\dagger (U^\dagger Y^\prime U) X,
\end{equation}
or
\begin{equation}
Y^{\prime \ast}
= (U^\ast X^\dagger U^\dagger) Y^\prime (U X U^\top)
= X^{\prime \dagger} Y^\prime X^\prime,
\end{equation}
as required.
A similar argument holds for the quartic terms and the proof is complete.

The fact that the transpose $U^\top$ appears in Eq.~(\ref{X-prime})
rather than $U^\dagger$ is crucial.
In Eq.~(\ref{S-prime}),
applicable to HF symmetries,
$U^\dagger$ appears.
Consequently,
a basis may be chosen where the HF symmetry is represented by
a diagonal matrix $S$.
The presence of $U^\top$ in Eq.~(\ref{X-prime}) implies
that, 
contrary to popular belief,
\textit{it is not possible to reduce all GCP transformations
to the standard CP transformation} of Eq.~(\ref{StandardCP})
by a basis transformation.
What is possible,
as we shall see below,
is to reduce an invariance of the THDM potential under any GCP
transformation,
to an invariance under the standard CP transformation
plus some extra constraints.

To be more specific, the following result is easily established. If
the unitary matrix $X$ is symmetric, then it follows
that\footnote{Here, we make use of a theorem in linear algebra that
states that for any unitary symmetric matrix $X$, a unitary matrix
$V$ exists such that $X=VV^\top$.  A proof of this result can be
found, e.g., in Appendix B of Ref.~\cite{GunHab}.}
a unitary matrix
$U$ exists such that $X'=UXU^\top=1$, in which case
$Y^{\prime\,*}=Y^\prime$.  In this case, a basis exists in which the
GCP is a standard CP transformation.
In contrast, if the unitary matrix $X$ is not symmetric, then no
basis exists in which $Y$ and $Z$ are real for generic values of the
scalar potential parameters.
Nevertheless, as we shall demonstrate below, by \textit{imposing}
the GCP symmetry on the scalar potential, the parameters of the
scalar potential are constrained in such a way that for an
appropriately chosen basis change, $Y^{\prime\,*}
=X^{\prime\,\dagger}Y'X'=Y'$ (with a similar result for $Z'$).

GCP transformations were studied in 
Refs.~\cite{GCP1,GCP2}. In
particular, Ecker, Grimus, and Neufeld \cite{GCP2} proved that for
every matrix $X$ there exists a unitary matrix $U$ such that
$X^\prime$ can be reduced to the form
\begin{equation}
X^\prime = UXU^\top=
\left(
\begin{array}{cc}
   \phantom{-}\cos{\theta} & \quad \sin{\theta}\\
   - \sin{\theta} & \quad \cos{\theta}
\end{array}
\right),
\label{GCP-reduced}
\end{equation}
where $0 \leq \theta \leq \pi/2$. Notice the restricted range for
$\theta$. The value of $\theta$ can be determined in either of two
ways: (i) the eigenvalues of $(X+X^\top)^\dagger(X+X^\top)/2$ are
$\cos{\theta}$, each of which is twice degenerate; or (ii) $X
X^\ast$ has the eigenvalues $e^{\pm 2 i \theta}$.

\subsection{The three classes of GCP symmetries}

Having reached the special form of $X^\prime$ in
Eq.~(\ref{GCP-reduced}),
we will now follow the strategy adopted by Ferreira and
Silva \cite{FS2} in connection with HF symmetries.
We substitute Eq.~(\ref{GCP-reduced}) for $X$ in
Eq.~(\ref{YZ-CPtransf}),
in order to identify the constraints imposed by this
reduced form of the GCP transformations on
the quadratic and quartic couplings.
For each value of $\theta$,
certain constraints will be forced upon the couplings.
If two different values of $\theta$ enforce the same constraints,
we will say that they are in the same class
(since no experimental distinction between the two will then be
possible).
We will start by considering the special cases of $\theta=0$
and $\theta=\pi/2$,
and then turn our attention to $0 < \theta < \pi/2$.

\subsubsection{CP1: $\theta=0$}

When $\theta=0$,
$X^\prime$ is the unit matrix,
and we obtain the standard CP transformation,
\begin{eqnarray}
\Phi_1 &\rightarrow& \Phi_1^\ast,
\nonumber\\
\Phi_2 &\rightarrow& \Phi_2^\ast, \label{eq:cp1}
\end{eqnarray}
under which Eqs.~(\ref{YZ-CPtransf}) take the very simple
form
\begin{eqnarray}
Y_{ab}^\ast
&=&
 Y_{ab} ,
\nonumber\\
Z_{ab,cd}^\ast
&=&
Z_{ab,cd}.
\end{eqnarray}
We denote this CP transformation by CP1.
It forces all couplings to be real.
Since most couplings are real by the Hermiticity of
the Higgs potential,
the only relevant constraints are
$\textrm{Im}\, m_{12}^2 = \textrm{Im}\, \lambda_5 =
\textrm{Im}\, \lambda_6 = \textrm{Im}\, \lambda_7 = 0$.

\subsubsection{CP2: $\theta=\pi/2$}

When $\theta=\pi/2$,
\begin{equation}
X^\prime =
\left(
\begin{array}{cc}
   \phantom{-} 0 & \quad 1\\
   - 1 & \quad 0
\end{array}
\right), \label{eq:cp2}
\end{equation}
and we obtain the CP transformation,
\begin{eqnarray}
\Phi_1 &\rightarrow& \Phi_2^\ast,
\nonumber\\
\Phi_2 &\rightarrow& - \Phi_1^\ast,
\end{eqnarray}
which we denote by CP2.
This was considered by Davidson and Haber \cite{DavHab}
in their Eq.~(37),
who noted that if this symmetry holds in one basis,
it holds in \textit{all} basis choices.
Under this transformation,
Eq.~(\ref{DY-GCP}) forces the matrix of quadratic
couplings to obey
\begin{equation}
0 =
\Delta Y =
\left(
\begin{array}{cc}
   m_{11}^2 - m_{22}^2 & \quad -2 m_{12}^2\\
   - 2 m_{12}^{2 \ast} & \quad m_{22}^2 - m_{11}^2,
\end{array}
\right)
\end{equation}
leading to $m_{22}^2 = m_{11}^2$ and $m_{12}^2=0$.
Similarly,
we may construct a matrix of matrices containing
all coefficients $\Delta Z_{ab,cd}$.
The uppermost-leftmost matrix corresponds to $\Delta Z_{11,cd}$.
The next matrix along the same line corresponds
to $\Delta Z_{12,cd}$, and so on.
To enforce invariance under CP2,
we equate it to zero,
\begin{equation}
0 =
\left(
\begin{array}{cc}
   \left(
   \begin{array}{cc}
      \lambda_1 - \lambda_2 & \quad
      \lambda_6 + \lambda_7 \\
      \lambda_6^\ast + \lambda_7^\ast & \quad
      0
   \end{array}
   \right)
   &
   \left(
   \begin{array}{cc}
      \lambda_6 + \lambda_7 & \quad
      0 \\
      0 & \quad
      \lambda_6 + \lambda_7
   \end{array}
   \right)
   \\*[7mm]
   \left(
   \begin{array}{cc}
   \lambda_6^\ast + \lambda_7^\ast & \quad
   0 \\
   0 & \quad
   \lambda_6^\ast + \lambda_7^\ast
   \end{array}
   \right)
   &
   \left(
   \begin{array}{cc}
   0 & \quad
   \lambda_6 + \lambda_7 \\
   \lambda_6^\ast + \lambda_7^\ast & \quad
   \lambda_2 - \lambda_1
   \end{array}
   \right)
\end{array}
\right).
\end{equation}
We learn that invariance under CP2 forces
$m_{22}^2 = m_{11}^2$ and $m_{12}^2=0$,
$\lambda_2=\lambda_1$, and $\lambda_7 = - \lambda_6$,
leading precisely to the ERPS of Eq.~(\ref{ERPS}).
Recall that Gunion and Haber \cite{GunHab} found that,
under these conditions we can always find a basis where
all parameters are real.
As a result,
if the potential is invariant under CP2,
there is a basis where CP2 still holds and in which
the potential is also invariant under CP1.

\subsubsection{CP3: $0 < \theta < \pi/2$}

Finally we turn to the cases where $0 < \theta < \pi/2$.
Imposing Eq.~(\ref{DY-GCP}) yields
\begin{eqnarray}
0 = \Delta Y_{11} &=&
\left[ (m_{11}^2 - m_{22}^2)\ s - 2\ \textrm{Re}\, m_{12}^2\ c \right] s,
\nonumber\\
0 = \Delta Y_{22} &=&
- \Delta Y_{11},
\nonumber\\
0 = \Delta Y_{12} &=&
\textrm{Re}\, m_{12}^2\ ( c_2 - 1) - 2 i\ \textrm{Im}\, m_{12}^2
       + \tfrac{1}{2} (m_{22}^2 - m_{11}^2)\ s_2,
\end{eqnarray}
where we have used $c=\cos{\theta}$, $s=\sin{\theta}$,
$c_2=\cos{2 \theta}$, and $s_2=\sin{2 \theta}$.
Since $\theta \neq 0, \pi/2$,
the conditions $m_{22}^2 = m_{11}^2$ and $m_{12}^2=0$ are imposed,
as in CP2.
Similarly, Eq.~(\ref{DZ-GCP}) yields
\begin{eqnarray}
0 = \Delta Z_{11,11} &=&
\lambda_1 (1-c^4) - \lambda_2 s^4
- \tfrac{1}{2} \lambda_{345} s_2^2
+ 4\ \textrm{Re}\, \lambda_6 c^3 s + 4\ \textrm{Re}\, \lambda_7 c s^3,
\nonumber\\
0 = \Delta Z_{22,22} &=&
\lambda_2 (1-c^4) - \lambda_1 s^4
- \tfrac{1}{2} \lambda_{345} s_2^2
- 4\ \textrm{Re}\, \lambda_7 c^3 s - 4\ \textrm{Re}\, \lambda_6 c s^3,
\nonumber\\
0 = \Delta Z_{11,22}
&=&
- \tfrac{1}{4} s_2
\left[
4 \textrm{Re}\, (\lambda_6 - \lambda_7) c_2
+ (\lambda_1 + \lambda_2 - 2 \lambda_{345})s_ 2
\right],
\nonumber\\
0 = \Delta Z_{12,21} &=&
\Delta Z_{11,22}
\nonumber\\
0 = \textrm{Re}\, \Delta Z_{11,12} &=&
\tfrac{1}{4} s \left[
(-3 \lambda_1 + \lambda_2 + 2 \lambda_{345}) c
- (\lambda_1 + \lambda_2 - 2 \lambda_{345}) c_3
\right.
\nonumber\\
& & \hspace{7mm}
\left.
+ 4 \textrm{Re}\, \lambda_6 (2 s + s_3)
- 4 \textrm{Re}\, \lambda_7 s_3
\right],
\nonumber\\
0 = \textrm{Re}\, \Delta Z_{22,12} &=&
\tfrac{1}{4} s \left[
(- \lambda_1 + 3 \lambda_2 - 2 \lambda_{345}) c
+ (\lambda_1 + \lambda_2 - 2 \lambda_{345}) c_3
\right.
\nonumber\\
& & \hspace{7mm}
\left.
- 4 \textrm{Re}\, \lambda_6 s_3
+ 4 \textrm{Re}\, \lambda_7 (2 s + s_3)
\right],
\nonumber\\
0 = \textrm{Re}\, \Delta Z_{12,12} &=&
\Delta Z_{11,22}
\label{Real-5}
\\
0 = \textrm{Im}\, \Delta Z_{11,12} &=&
\tfrac{1}{2}
\left[
\textrm{Im}\, \lambda_6 (3+c_2)
+ \textrm{Im}\, \lambda_7 (1-c_2)
- \textrm{Im}\, \lambda_5 s_2
\right],
\nonumber\\
0 = \textrm{Im}\, \Delta Z_{22,12} &=&
\tfrac{1}{2}
\left[
\textrm{Im}\, \lambda_6 (1-c_2)
+ \textrm{Im}\, \lambda_7 (3+c_2)
+ \textrm{Im}\, \lambda_5 s_2
\right],
\nonumber\\
0 = \textrm{Im}\, \Delta Z_{12,12} &=&
2 c
\left[
\textrm{Im}\, \lambda_5 c + \textrm{Im}\,(\lambda_6-\lambda_7)s
\right],
\label{Im-3}
\end{eqnarray}
where $\lambda_{345}=\lambda_3 + \lambda_4 + \textrm{Re}\, \lambda_5$,
$c_3 = \cos{3 \theta}$, and $s_3 = \sin{3 \theta}$.

The last three equations may be written as
\begin{equation}
0=
\left[
\begin{array}{ccc}
-s_2 &\quad (3+c_2) & \quad (1-c_2)\\
\phantom{-}s_2 &\quad  (1-c_2) &\quad  (3+c_2)\\
(1+c_2) &\quad  s_2 &\quad  -s_2
\end{array}
\right]
\left[
\begin{array}{c}
\textrm{Im}\, \lambda_5 \\
\textrm{Im}\, \lambda_6 \\
\textrm{Im}\, \lambda_7
\end{array}
\right].
\end{equation}
The determinant of this homogeneous system of three equations
in three unknowns is $32 c^2$,
which can never be zero since we are assuming that $\theta \neq \pi/2$.
As a result,
$\lambda_5$, $\lambda_6$, and $\lambda_7$ are real,
whatever the value of $0 < \theta < \pi/2$ chosen for the GCP
transformation.
Since $m_{12}^2=0$,
all potentially complex parameters must be real.
We conclude that a potential invariant
under any GCP with $0 < \theta < \pi/2$
is automatically invariant under CP1.
Combining this with what we learned from CP2,
we conclude the following:
if a potential is invariant under some GCP transformation,
then a basis may be found in which it is also invariant
under the standard CP transformation,
with some added constraints on the parameters.

The other set of five independent homogeneous equations in
five unknowns has a determinant equal to zero,
meaning that not all parameters must vanish.
We find that
\begin{eqnarray}
0=
\Delta Z_{11,11} - \Delta Z_{22,22}
&=&
2s \left[
s\ (\lambda_1-\lambda_2) + c\ 2 \textrm{Re}\, (\lambda_6 + \lambda_7)
\right],
\nonumber\\
0=
\textrm{Re}\, \Delta Z_{11,12}
- \textrm{Re}\, \Delta Z_{22,12}
&=&
s \left[
-c\ (\lambda_1-\lambda_2) + s\ 2 \textrm{Re}\, (\lambda_6 + \lambda_7)
\right].
\end{eqnarray}
Since $s \neq 0$,
we obtain the homogeneous system
\begin{equation}
0=
\left[
\begin{array}{cc}
\phantom{-}s & \quad c\\
-c & \quad s
\end{array}
\right]
\left[
\begin{array}{c}
\lambda_1 - \lambda_2 \\
2 \textrm{Re}\, (\lambda_6 + \lambda_7)
\end{array}
\right],
\end{equation}
whose determinant is unity.
We conclude that $\lambda_2 = \lambda_1$ and $\lambda_7 = - \lambda_6$.
Thus,
GCP invariance with any value of $0 < \theta \leq \pi/2$ leads
to the ERPS of Eq.~(\ref{ERPS}).
Substituting back we obtain
$\Delta Z_{11,11} = \Delta Z_{22,22} = - \Delta Z_{11,22}$
and
$\textrm{Re}\, \Delta Z_{11,12} = - \textrm{Re}\, \Delta Z_{22,12}$,
leaving only two independent equations:
\begin{eqnarray}
0=
\Delta Z_{11,11}
&=&
\tfrac{1}{2} s_2 \left[
(\lambda_1 - \lambda_{345}) s_2 + 4 \lambda_6 c_2 \right],
\nonumber\\
0=
\textrm{Re}\, \Delta Z_{22,12}
&=&
\tfrac{1}{2} s_2 \left[
(\lambda_1 - \lambda_{345}) c_2 - 4 \lambda_6 s_2 \right],
\end{eqnarray}
where we have used $c + c_3 = 2 c c_2$ and $s + s_3 = 2 c s_2$.
Since $s_2 \neq 0$,
the determinant of the system does not vanish,
forcing $\lambda_1=\lambda_{345}$ and $\lambda_6=0$.

Notice that our results do not depend on which exact
value of $ 0 < \theta < \pi/2$ in Eq.~(\ref{GCP-reduced}) we have chosen.
If we require invariance of the potential under GCP with some
particular value of $ 0 < \theta < \pi/2$,
then the potential is immediately invariant under GCP
with any other value of $ 0 < \theta < \pi/2$.
We name this class of CP invariances, CP3.
Combining everything,
we conclude that invariance under CP3 implies
\begin{eqnarray}
m_{11}^2 = m_{22}^2,
& \hspace{4ex} &
m_{12}^2=0,
\nonumber\\
\lambda_2 = \lambda_1,
& \hspace{4ex} &
\lambda_7 = \lambda_6 = 0,
\nonumber\\
\textrm{Im}\, \lambda_5 = 0,
& \hspace{4ex} &
\textrm{Re}\, \lambda_5 = \lambda_1 - \lambda_3 -\lambda_4.
\label{Region-CP3}
\end{eqnarray}
The results of this section are all summarized in Table~\ref{master1}
of section~\ref{sec:summary}.

\subsection{The square of the GCP transformation}

If we apply a GCP transformation twice to the scalar fields, we will
have, from Eq.~\eqref{GCP}, that
\begin{equation}
\left(\Phi^{\textrm{GCP}}_a\right)^{\textrm{GCP}}
\;=\;
X_{a \alpha} \left(\Phi^{\textrm{GCP}}_\alpha\right)^\ast
\;=\;
X_{a \alpha}\, X_{\alpha b}^\ast\ \Phi_b \;\;\; ,
\end{equation}
so that the square of a GCP transformation is given by
\begin{equation}
(GCP)^2 \;=\; XX^\ast \;\;\; .
\label{eq:cpq}
\end{equation}
In particular, for a generic unitary matrix $X$, $(GCP)^2$ is a Higgs Family 
symmetry transformation.

Usually, only GCP transformations with $(GCP)^2 = \boldsymbol{1}$
(where $\boldsymbol{1}$ is the unit matrix)
are considered in the literature.
For such a situation,
$X=X^\dagger=X^*$,
and one can always find
a basis in which $X=\boldsymbol{1}$.
In this case,
a GCP transformation is equivalent to a standard CP
transformation in the latter basis choice.
For example,
the restriction that $(GCP)^2 = \boldsymbol{1}$
(or equivalently, requiring the squared of the corresponding generalized 
time-reversal transformation to equal the unit matrix)
was imposed in 
Ref.~\cite{GunHab} and more recently in Ref.~\cite{mani}. 
However, as we have illustrated in this section, the invariance
under a GCP transformation, in which  $(GCP)^2 \neq \boldsymbol{1}$ 
(corresponding to a unitary matrix $X$ that is not symmetric)
is a \textit{stronger} restriction on the parameters of the 
scalar potential than the invariance under a standard CP transformation.   

As we see from the results in the previous sections,
$X$ is {\em not} symmetric for the symmetries CP2 and CP3.
In fact, this feature provides a strong distinction among the
three GCP symmetries previously introduced.  
Let us briefly examine $(GCP)^2$ for the three possible
cases $CP1$, $CP2$ and $CP3$.

\subsubsection{$(CP1)^2$}

Comparing Eqs.~\eqref{GCP} and~\eqref{eq:cp1}, we come to the
immediate conclusion that $X_{CP1}\,=\,\boldsymbol{1}$, so that
Eq.~\eqref{eq:cpq} yields
\begin{equation}
(CP1)^2\;=\;\boldsymbol{1}\,.
\end{equation}
This implies that a CP1-invariant scalar potential 
is invariant under the symmetry group
$Z_2=\{\boldsymbol{1}\,,\,CP1\}$.

\subsubsection{$(CP2)^2$}

The matrix $X_{\textrm{CP2}}$ is shown in Eq.~\eqref{eq:cp2} so that, by
Eq.~\eqref{eq:cpq}, we obtain
\begin{equation}
(CP2)^2\;=\;-\,\boldsymbol{1}\,.
\end{equation}
Although this result significantly distinguished CP2 from CP1, 
the authors of Ref.~\cite{mani} noted (in considering their
$CP_g^{(i)}$ symmetries) that the transformation law for $\Phi_a$
under (CP2)$^2$ can be reduced to the identity by a global
hypercharge transformation.  That is,
if we start with the symmetry group $Z_4=\{\boldsymbol{1}\,,\,CP2\,,\,
-\boldsymbol{1}\,,\,-CP2\}$, we can impose an equivalence relation
by identifying two elements of $Z_4$ related by multiplication  
by $-\boldsymbol{1}$.  If we denote $(Z_2)_Y=\{\boldsymbol{1}\,,
-\boldsymbol{1}\}$ as the two-element
discrete subgroup of the global hypercharge
$U(1)_Y$, then the discrete symmetry group that is orthogonal to $U(1)_Y$
is given by $Z_4/(Z_2)_Y\iso Z_2$.   Hence,
the CP2-invariant scalar potential exhibits
a $Z_2$ symmetry orthogonal to the Higgs flavor symmetries
of the potential.

\subsubsection{$(CP3)^2$}

The matrix $X_{\textrm{CP3}}$ is given in Eq.~\eqref{GCP-reduced}, with
$0<\theta<\pi/2$, so that, by Eq.~\eqref{eq:cpq}, we obtain
\begin{equation}
(CP3)^2\;=\;\left(
\begin{array}{cc}
  \phantom{-} \cos{2\theta} & \quad \sin{2\theta}\\
   - \sin{2\theta} & \quad \cos{2\theta}
\end{array}
\right)\,,
\end{equation}
which once again is {\em not} the unit matrix.
However, the transformation law for $\Phi_a$ under (CP3)$^2$ 
\textit{cannot} be reduced to the identity by a global
hypercharge transformation. 
This is the reason why Ref.~\cite{mani} did not consider CP3.
However, $(CP3)^2$ is a non-trivial HF symmetry of the CP3-invariant
scalar potential.\footnote{In Section~\ref{sec:summary}B, we shall
identify $(CP3)^2$ with the Peccei Quinn U(1) symmetry defined as
in Eq.~(\ref{U1}) and then transformed to a new basis
according to the unitary matrix defined in Eq.~(\ref{UPQ}).}
Thus, one can always reduce the square of 
CP3 to the identity by applying a suitable HF symmetry transformation.
In particular, a CP3-invariant scalar potential also exhibits a $Z_2$ symmetry
that is orthogonal to the Higgs flavor symmetries of the potential.

In this paper, we prove that there are three and only three
classes of GCP transformations.
Of course, within each class,
one may change the explicit form of the
scalar potential by a suitable basis transformation;
but that will not alter its physical consequences.
Similarly,
one can set some parameters to zero in some ad-hoc fashion,
not rooted in a symmetry requirement. 
But, as we have shown, the constraints imposed on the scalar potential
by a single GCP symmetry can be grouped into three classes:
CP1, CP2, and CP3.

\section{\label{sec:summary}Classification of the HF and
GCP transformation classes in the THDM}

\subsection{Constraints on scalar potential parameters}

Suppose that one is allowed one single symmetry requirement
for the potential in the THDM.
One can choose an invariance under one particular Higgs Family
symmetry.
We know that there are only two independent classes
of such simple symmetries: $Z_2$ and Peccei-Quinn $U(1)$.
One can also choose an invariance under a particular
GCP symmetry.  
We have proved that there are three classes of
GCP symmetries, named CP1, CP2, and CP3.
If any of the above symmetries is imposed on the THDM scalar
potential (in a specified basis), then the coefficients
of the scalar potential are constrained, as summarized in
Table~\ref{master1}.  For completeness, we also exhibit
the constraints imposed by $SO(3)$,
the largest possible continuous HF symmetry that is orthogonal to
the global hypercharge $U(1)_Y$ transformation.
%
\begin{table}[ht!]
\caption{Impact of the symmetries on the coefficients
of the Higgs potential in a specified basis.}
\begin{ruledtabular}
\begin{tabular}{ccccccccccc}
symmetry & $m_{11}^2$ & $m_{22}^2$ & $m_{12}^2$ &
$\lambda_1$ & $\lambda_2$ & $\lambda_3$ & $\lambda_4$ &
$\lambda_5$ & $\lambda_6$ & $\lambda_7$ \\
\hline
$Z_2$ &   &   & 0 &
   &  &  &  &
   & 0 & 0 \\
$U(1)$ &  &  & 0 &
 &  & &  &
0 & 0 & 0 \\
$SO(3)$ &  & $ m_{11}^2$ & 0 &
   & $\lambda_1$ &  & $\lambda_1 - \lambda_3$ &
0 & 0 & 0 \\
\hline
$\Pi_2$ &  & $ m_{11}^2$ & real &
   & $ \lambda_1$ &  &  &
real &  & $\lambda_6^\ast$
\\
\hline
CP1 &  &  & real &
 & &  &  &
real & real & real \\
CP2 &  & $m_{11}^2$ & 0 &
  & $\lambda_1$ &  &  &
   &  & $- \lambda_6$ \\
CP3 &  & $m_{11}^2$ & 0 &
   & $\lambda_1$ &  &  &
$\lambda_1 - \lambda_3 - \lambda_4$ (real) & 0 & 0 \\
\end{tabular}
\end{ruledtabular}
\label{master1}
\end{table}
%

Empty entries in Table~\ref{master1} correspond to a lack of constraints on the
corresponding parameters.
Table~\ref{master1}has been constructed for those basis choices in which
$Z_2$ and $U(1)$ have the specific forms in Eqs.~(\ref{Z2}) and
(\ref{U1}), respectively.
If, for example,
the basis is changed and $Z_2$ acquires the form $\Pi_2$ in
Eqs.~(\ref{Pi2}),
then the constraints on the coefficients are altered,
as shown explicitly on the
fourth line of Table~\ref{master1}.
However,
this does not correspond to a new model.
All physical predictions are the same since
the specific forms of $Z_2$ and $\Pi_2$ differ only by
the basis change in Eq.~(\ref{Z2ToPi2}).
The constraints for CP1, CP2, and CP3 shown in Table I
apply to the basis in which the GCP transformation
of Eq.~(\ref{GCP}) is used where $X$ has been transformed into $X^\prime$
given by Eq.~(\ref{GCP-reduced}),
with $\theta=0$, $\theta = \pi/2$,
and $0 < \theta < \pi/2$, respectively.

\subsection{Multiple symmetries and GCP}

We now wish to consider the possibility of simultaneously imposing
more than one symmetry requirement on the Higgs potential.
For example, one can require that $Z_2$ and $\Pi_2$ be enforced
\textit{within the same basis}.  In what follows, we shall indicate
that the two symmetries are enforced simultaneously by writing
$Z_2\oplus\Pi_2$.
Combining the constraints from the appropriate
rows of Table~\ref{master1},
we conclude that,
under these two simultaneous requirements
\begin{eqnarray}
m_{22}^2 = m_{11}^2,
& \hspace{4ex} &
m_{12}^2=0,
\nonumber\\
\lambda_2 = \lambda_1,
& \hspace{4ex} &
\lambda_7 = \lambda_6 = 0,
\hspace{4ex} \textrm{Im}\, \lambda_5 = 0.
\label{Z2+Pi2}
\end{eqnarray}
This coincides exactly with the conditions of the ERPS
in a very special basis,
as shown in Eq.~(\ref{ERPS2}).
Since CP2 leads to the ERPS of Eq.~(\ref{ERPS}),
we conclude that
\begin{equation}
Z_2 \oplus \Pi_2 \equiv \textrm{CP2 in some specific basis}.
\label{equiv-CP2}
\end{equation}
This was noted previously by Davidson and Haber \cite{DavHab}.
Now that we know what all classes of HF and CP symmetries can
look like,
we can ask whether all GCP symmetries can be written as
the result of some multiple HF symmetry.

This is clearly not possible for CP1 because of parameter counting.
Table~\ref{master1} shows that CP1 reduces the scalar potential to ten
real parameters.
We can still perform an orthogonal basis change while keeping
all parameters real.
This freedom can be used to remove one further parameter;
for example, setting $m_{12}^2=0$ by diagonalizing the $Y$
matrix.
No further simplification is allowed.
As a result, CP1 leaves nine independent parameters.
The smallest HF symmetry is $Z_2$.
Table~\ref{master1} shows that $Z_2$ reduces the potential to six
real and one complex parameter.
The resulting eight parameters could never account for the
nine needed to fully describe the most general model
with the standard CP invariance CP1.\footnote{In
Ivanov's language, this is clear since CP1 corresponds to
a $Z_2$ transformation of the vector $\vec{r}$,
which is the simplest transformation on $\vec{r}$ one
could possibly make.
See section~\ref{more_multiple}.}

But one can utilize two HF symmetries in order
to obtain the same constraints obtained by invariance under CP3.
Let us impose \textit{both} $U(1)$ and $\Pi_2$
\textit{in the same basis}.
From Table~\ref{master1},
we conclude that,
under these two simultaneous requirements
\begin{eqnarray}
m_{22}^2 = m_{11}^2,
& \hspace{4ex} &
m_{12}^2=0,
\nonumber\\
\lambda_2 = \lambda_1,
& \hspace{4ex} &
\lambda_7 = \lambda_6 = 0,
\hspace{4ex} \lambda_5 = 0.
\label{U1+Pi2}
\end{eqnarray}
This does not coincide with the
conditions for invariance under CP3 shown in Eq.~(\ref{Region-CP3}).
However,
one can use the transformation rules in Eqs.~(A13)-(A23)
of Davidson and Haber \cite{DavHab},
in order to show that a basis transformation, 
\begin{equation} \label{UPQ}
U=\frac{1}{\sqrt{2}}\left(\begin{array}{cc} \phantom{-}1 & \quad -i \\
-i & \quad\phantom{-}1\end{array}\right)\,, 
\end{equation}
may be chosen which takes us from Eqs.~(\ref{Region-CP3}),
where $\textrm{Re}\, \lambda_5 = \lambda_1 -\lambda_3 -\lambda_4$,
to Eqs.~(\ref{U1+Pi2}),
where $\lambda_5=0$ (while maintaining the other relations among
the scalar potential parameters).
We conclude that
\begin{equation}
U(1) \oplus \Pi_2 \equiv \textrm{CP3 in some specific basis}.
\label{equiv-CP3}
\end{equation}
Note that in the basis in which the CP3 relations of  Eq.~(\ref{Region-CP3})
are satisfied with $\lambda_5\neq 0$, the discrete HF symmetry
$\Pi_2$ is still respected.
However, using Eq.~(\ref{UPQ}), it follows that the U(1)-Peccei Quinn
symmetry corresponds to the invariance of the scalar potential under
$\Phi_a\to \mathcal{O}_{ab}\Phi_b$, where $\mathcal{O}$ is an arbitrary
$SO(2)$ matrix.

The above results suggest that it should be possible to distinguish CP1,
CP2, and CP3 in a basis invariant fashion.
Botella and Silva \cite{BS} have built three so-called $J$-invariants
that detect any signal of CP violation (either explicit or
spontaneous) after the minimization
of the scalar potential.  However, in this paper we are concerned
about the symmetries of the scalar potential independently of the
choice of vacuum.  Thus, we shall consider
the four so-called $I$-invariants built by
Gunion and Haber \cite{GunHab} in order to detect any
signal of \textit{explicit} CP violation present (before the vacuum state
is determined).
If any of these invariants is nonzero, then CP is explicitly violated,
and neither CP1, nor CP2, nor CP3 hold.
Conversely,
if all $I$-invariants are zero, then CP is explicitly conserved, but we cannot
tell a priori which GCP applies.
Eqs.~(\ref{equiv-CP2}) and (\ref{equiv-CP3}) provide the crucial hint.
If we have CP conservation, $Z_2\oplus\Pi_2$ holds,
and $U(1)$ does not,
then we have CP2.
Alternatively,
if we have CP conservation, and $U(1)\oplus\Pi_2$ also holds,
then we have CP3.
We recall that both CP2 and CP3 lead to the ERPS,
and that the general conditions for the ERPS in Eq.~(\ref{ERPS})
are basis independent.
This allows us to distinguish CP2 and CP3 from CP1.
But, prior to the present work,
no basis-independent quantity had been identified in the literature
that could distinguish $Z_2$ and $U(1)$ in the ERPS.
The basis-independent quantity $D$ introduced
in subsection~\ref{subsec:D} is precisely the invariant required for
this task.  That is,
in the ERPS $D\neq0$ implies CP2, whereas $D=0$ implies CP3.

One further consequence of the results of Table~\ref{master1}
can be seen by simultaneously imposing the U(1) Peccei-Quinn symmetry
and the CP3 symmetry \textit{in the same basis}.  The resulting
constraints on the scalar potential parameters are precisely those of
the SO(3) HF symmetry.  Thus, we conclude that
\begin{equation}
U(1) \oplus \textrm{CP3} \equiv SO(3).
\label{equiv-O3}
\end{equation}
In particular, $SO(3)$ is not a simple HF symmetry, as the invariance
of the scalar potential under a single element of SO(3) is not
sufficient to guarantee invariance under the full SO(3) group of
transformations.

\subsection{Maximal symmetry group of the scalar potential
orthogonal to $U(1)_Y$}

The standard CP symmetry, CP1,
is a discrete $Z_2$ symmetry that transforms the scalar
fields into their complex conjugates, and hence 
is not a subgroup of the $U(2)$ transformation group
of Eq.~\ref{basis-transf}.  We have previously noted that
THDM scalar potentials that exhibit \textit{any} non-trivial
HF symmetry $G$ is automatically CP-conserving.  Thus, the actual
symmetry group of the scalar potential is in fact 
the semidirect product\footnote{In general, the non-trivial element of
$Z_2$ will not commute with all elements of $G$, in which case the
relevant mathematical structure is that of a semidirect product.  In
cases where the non-trivial element of $Z_2$ commutes with all
elements of $G$, we denote the corresponding direct product as
$G\otimes Z_2$.}
of $G$ and $Z_2$, which we write as $G\rtimes Z_2$.
Noting that $U(1)\rtimes Z_2\iso SO(2)\rtimes Z_2\iso O(2)$, and
$SO(3)\otimes Z_2\iso O(3)$, we conclude that the maximal 
symmetry groups of the scalar potential orthogonal to $U(1)_Y$
for the possible choices of HF symmetries are given in 
Table~\ref{maximal}.\footnote{For ease of notation, we denote
$Z_2\otimes Z_2$ by $(Z_2)^2$ and $Z_2\otimes Z_2\otimes Z_2$
by $(Z_2)^3$.}

\begin{table}[ht]
\caption{Maximal symmetry groups [orthogonal to global $U(1)_Y$
hypercharge] of the scalar sector of the THDM.}
\begin{ruledtabular}
\begin{tabular}{ccc}
designation &  HF symmetry group  & maximal symmetry group\\
\hline
$Z_2$ & $Z_2$ & $(Z_2)^2$ \\
Peccei-Quinn & $U(1)$ & $O(2)$ \\
$SO(3)$ & $SO(3)$ & $O(3)$ \\
CP1  & --- & $Z_2$ \\
CP2  & $(Z_2)^2$ & $(Z_2)^3$ \\
CP3 & $O(2)$ & $O(2)\otimes Z_2$
\\
\end{tabular}
\end{ruledtabular}
\label{maximal}
\end{table}
%

Finally, we reconsider CP2 and CP3.  Eq.~(\ref{equiv-CP2})
implies that the CP2 symmetry is equivalent to a $(Z_2)^2$ HF
symmetry.  To prove this statement, we note that in the
two-dimensional flavor space of Higgs fields, the $Z_2$ and $\Pi_2$
discrete symmetries defined by Eqs.~(\ref{Z2}) and (\ref{Pi2})
are given by:
\begin{equation}
Z_2=\{S_0\,,\,S_1\}\,,\qquad\qquad \Pi_2=\{S_0\,,\,S_2\}\,,
\end{equation}
where $S_0\equiv \boldsymbol{1}$ is the $2\times 2$ identity matrix and
\begin{equation}
S_1=\left(
\begin{array}{cc}
1 & \quad\phantom{-} 0 \\
0 & \quad -1 \\
\end{array}
\right)\,,\qquad\qquad
S_2=\left(
\begin{array}{cc}
0 & \quad 1 \\
1 & \quad 0 \\
\end{array}
\right)\,.
\end{equation}
If we impose the $Z_2$ and $\Pi_2$ symmetry in the same basis, then the
scalar potential is invariant under the dihedral group of eight elements,
\begin{equation}
D_4=\{S_0\,,\,S_1\,,\,S_2\,,\,S_3\,,\,-S_0\,,\,-S_1\,,\,-S_2\,,\,-S_3\}\,,
\end{equation}
where $S_3=S_1 S_2=-S_2 S_1$.  
As before, we identify $(Z_2)_Y\equiv\{S_0\,,\,-S_0\}$ as the
two-element discrete subgroup of the global hypercharge $U(1)_Y$.
However, we have defined the HF symmetries to be orthogonal to $U(1)_Y$.
Thus, to determine the HF symmetry group of CP2, we identify as
equivalent those elements of $D_4$ that are related by multiplication
by $-S_0$.  Group theoretically, we identify the HF symmetry group
of CP2 as
\begin{equation}
D_4/(Z_2)_Y\iso Z_2\otimes Z_2\,.
\end{equation}

The HF symmetry group of CP2 is not the maximally allowed symmetry
group.  In particular, the constraints of CP2 on the scalar potential
imply the existence of a basis in which all scalar potential
parameters are real.  Thus, the scalar potential is explicitly
CP-conserving.  The $Z_2$ symmetry associated with this CP
transformation is orthogonal to the HF symmetry as previously noted.  
(This is easily checked explicitly by employing a four-dimensional real
representation of the two complex scalar fields.)  Thus,
the maximal symmetry group of the CP2-symmetric scalar potential
is $(Z_2)^3$.  Similarly, Eq.~(\ref{equiv-CP3})
implies that the CP3 symmetry is equivalent to a $U(1)\rtimes Z_2$ HF
symmetry. This is isomorphic to an $O(2)$ HF symmetry, which is
a subgroup of the maximally allowed $SO(3)$ HF symmetry group. 
However, the constraints of CP3 on the scalar potential
imply the existence of a basis in which all scalar potential
parameters are real.  Thus, the scalar potential is explicitly
CP-conserving.  Once again, the $Z_2$ symmetry associated with this CP
transformation is orthogonal to the HF symmetry noted above.  Thus,
the maximal symmetry group of the CP3-symmetric scalar potential
is $O(2)\otimes Z_2$.  

The above results are also summarized in 
Table~\ref{maximal}.  In all cases, the maximal symmetry group is
a direct product of the HF symmetry group and the $Z_2$ corresponding
to the standard CP-transformation, whose square is the identity operator.

One may now ask whether Table~\ref{maximal} exhausts all
possible independent symmetry constraints that one
may place on the Higgs potential.
Perhaps one can choose other combinations,
or maybe one can combine three, four, or more
symmetries.
We know of no way to answer this problem
based only on the transformations of the scalar
fields $\Phi_a$.
Fortunately,
Ivanov has solved this problem~\cite{Ivanov1} by looking
at the transformation properties of field bilinears,
thus obtaining for the first time the list of symmetries given in the last
column of Table~\ref{maximal}.

\subsection{\label{more_multiple}More on multiple symmetries}

We start by looking at the implications of the symmetries we have
studied so far on the vector $\vec{r} = \{ r_1, r_2, r_3\}$,
whose components were introduced in Eq.~(\ref{r_Ivanov}).
Notice that a unitary transformation $U$ on
the fields $\Phi_a$ induces an orthogonal
transformation $O$ on the vector of bilinears
$\vec{r}$,
given by Eq.~(\ref{O}).
For every pair of unitary transformations $\pm U$
of $SU(2)$,
one can find some corresponding transformation $O$
of $SO(3)$,
in a two-to-one correspondence.
We then see what these symmetries imply
for the coefficients of Eq.~(\ref{VH3})
(recall the $\Lambda_{\mu \nu}$ is a symmetric matrix).
Below, we list the transformation of $\vec{r}$ under which the
scalar potential is invariant, followed by the 
corresponding constraints on the
quadratic and quartic scalar potential parameters, $M_\mu$ and
$\Lambda_{\mu\nu}$.  

Using the results of Table I, we find that $Z_2$ implies
\begin{equation}
\vec{r} \rightarrow
\left[
\begin{array}{c}
-r_1\\
-r_2\\
\phantom{-}r_3
\end{array}
\right],
\hspace{10mm}
\left[
\begin{array}{c}
M_0\\
0\\
0\\
M_3
\end{array}
\right],
\hspace{4ex}
\left[
\begin{array}{cccc}
\Lambda_{00} &\,\,\, 0 &\,\,\, 0 &\,\,\, \Lambda_{03}\\
 0 &\,\,\, \Lambda_{11} &\,\,\, \Lambda_{12} &\,\,\, 0\\
 0 & \,\,\,\Lambda_{12} &\,\,\, \Lambda_{22} &\,\,\, 0\\
\Lambda_{03} &\,\,\, 0 &\,\,\, 0 &\,\,\, \Lambda_{33}
\end{array}
\right],
\label{Iv-Z2}
\end{equation}
$U(1)$ implies
\begin{equation}
\vec{r} \rightarrow
\left[
\begin{array}{ccc}
c_2 & -s_2 & \phantom{-}0\\
s_2 &\phantom{-}c_2 & \phantom{-}0\\
0 & \phantom{-}0 & \phantom{-}1
\end{array}
\right]\ \vec{r},
\hspace{10mm}
\left[
\begin{array}{c}
M_0\\
0\\
0\\
M_3
\end{array}
\right],
\hspace{4ex}
\left[
\begin{array}{cccc}
\Lambda_{00} &\,\,\, 0 & \,\,\,0 & \,\,\,\Lambda_{03}\\
 0 &\,\,\, \Lambda_{11} &\,\,\, 0 &\,\,\, 0\\
 0 & \,\,\,0 & \,\,\,\Lambda_{11} & \,\,\,0\\
\Lambda_{03} & \,\,\,0 & \,\,\,0 &\,\,\, \Lambda_{33}
\end{array}
\right],
\label{Iv-U1}
\end{equation}
and SO(3) implies 
\begin{equation}
\vec{r} \rightarrow
\mathcal{O} \vec{r},
\hspace{10mm}
\left[
\begin{array}{c}
M_0\\
0\\
0\\
0
\end{array}
\right],
\hspace{4ex}
\left[
\begin{array}{cccc}
\Lambda_{00} &\,\,\, 0 & \,\,\,0 & \,\,\,0\\
 0 &\,\,\, \Lambda_{11} &\,\,\, 0 &\,\,\, 0\\
 0 & \,\,\,0 & \,\,\,\Lambda_{11} & \,\,\,0\\
0 & \,\,\,0 & \,\,\,0 &\,\,\, \Lambda_{11}
\end{array}
\right],
\label{Iv-SO3}
\end{equation}
where $\mathcal{O}$ is an arbitrary $3\times 3$ orthogonal 
matrix of unit determinant.
In the language of bilinears, a basis invariant condition for the presence of
$SO(3)$ is that the three eigenvalues of $\tilde{\Lambda}$ are equal.
(Recall that $\tilde{\Lambda} = \left\{\Lambda_{ij}\right\}$; $i,j=1,2,3$).

As for the GCP symmetries,
CP1 implies
\begin{equation}
\vec{r} \rightarrow
\left[
\begin{array}{c}
\phantom{-}r_1\\
-r_2\\
\phantom{-}r_3
\end{array}
\right],
\hspace{10mm}
\left[
\begin{array}{c}
M_0\\
M_1\\
0\\
M_3
\end{array}
\right],
\hspace{4ex}
\left[
\begin{array}{cccc}
\Lambda_{00} & \,\,\,\Lambda_{01}  & \,\,\, 0 & \,\,\, \Lambda_{03}\\
\Lambda_{01} & \,\,\, \Lambda_{11} & \,\,\, 0 & \,\,\, \Lambda_{13}\\
 0 & \,\,\, 0 & \,\,\, \Lambda_{22} &  \,\,\,0\\
\Lambda_{03} &  \,\,\,\Lambda_{13} &  \,\,\,0 & \,\,\, \Lambda_{33}
\end{array}
\right],
\label{Iv-CP1}
\end{equation}
CP2 implies
\begin{equation}
\vec{r} \rightarrow
\left[
\begin{array}{c}
-r_1\\
-r_2\\
-r_3
\end{array}
\right],
\hspace{10mm}
\left[
\begin{array}{c}
M_0\\
0\\
0\\
0
\end{array}
\right],
\hspace{4ex}
\left[
\begin{array}{cccc}
\Lambda_{00} &\,\,\, 0 &\,\,\, 0 &\,\,\, 0\\
0 &\,\,\, \Lambda_{11}\,\,\, &\,\,\, \Lambda_{12} & \,\,\,\Lambda_{13}\\
0 &\,\,\, \Lambda_{12} &\,\,\, \Lambda_{22} &\,\,\, \Lambda_{23}\\
0 &\,\,\, \Lambda_{13} & \,\,\,\Lambda_{23} & \,\,\,\Lambda_{33}
\end{array}
\right],
\label{Iv-CP2}
\end{equation}
and CP3 implies
\begin{equation}
\vec{r} \rightarrow
\left[
\begin{array}{ccc}
\phantom{-}c_2 & \phantom{-}0 & \phantom{-}s_2\\
\phantom{-}0 & -1 & \phantom{-}0\\
-s_2 & \phantom{-}0 & \phantom{-}c_2
\end{array}
\right]\ \vec{r},
\hspace{10mm}
\left[
\begin{array}{c}
M_0\\
0\\
0\\
0
\end{array}
\right],
\hspace{4ex}
\left[
\begin{array}{cccc}
\Lambda_{00} &\,\,\, 0 & \,\,\,0 & 0\,\,\,\\
0 &\,\,\, \Lambda_{11} &\,\,\, 0 &\,\,\, 0\\
0 &\,\,\, 0 &\,\,\, \Lambda_{22} &\,\,\, 0\\
0 &\,\,\, 0 &\,\,\, 0 & \,\,\,\Lambda_{11}
\end{array}
\right].
\label{Iv-CP3}
\end{equation}
Notice that in CP3 two of the eigenvalues of $\Lambda$ are equal,
in accordance with our observation that $D$ can be used
to distinguish between CP2 and CP3.

Because each unitary transformation on the fields $\Phi_a$
induces an $SO(3)$ transformation on the vector
of bilinears $\vec{r}$,
and because the standard CP transformation
corresponds to an inversion of $r_2$
(a $Z_2$ transformation on the vector $\vec{r}$),
Ivanov \cite{Ivanov1}
considers all possible proper and improper transformations
of $O(3)$ acting on $\vec{r}$.  
He identifies the following six classes of transformations:
(i) $Z_2$; (ii) $(Z_2)^2$; (iii) $(Z_2)^3$;
(iv) $O(2)$; (v) $O(2) \otimes Z_2$; and (vi) $O(3)$.
Note that these symmetries are all orthogonal to the global $U(1)_Y$
hypercharge symmetry, as the bilinears $r_0$ and $\vec{r}$
are all singlets under a $U(1)_Y$ transformation.
The six classes above identified by Ivanov
correspond precisely to the six possible maximal
symmetry groups identified in Table~\ref{maximal}.
No other independent symmetry transformations are possible.

Our work permits one to identify the abstract transformation
of field bilinears utilized by Ivanov in terms of
transformations on the scalar fields themselves,
as needed for model building.
Combining our work with Ivanov's,
we conclude that there is only one new type
of symmetry requirement which one can place on
the Higgs potential via multiple symmetries.
Combining this with our earlier results,
we conclude that all possible symmetries on the scalar
sector of the THDM can be reduced to multiple HF symmetries,
with the exception of the standard CP transformation (CP1).

\section{\label{sec:allisCP}Building all symmetries with the standard CP}

We have seen that there are only six independent
symmetry requirements, listed in Table~\ref{maximal},
that  one can impose on the Higgs potential.
We have shown that all possible symmetries of the scalar
sector of the THDM can be reduced to multiple HF symmetries,
with the exception of the standard CP transformation (CP1).
Now we wish to show a dramatic result:
\textit{all possible symmetries on the scalar
sector of the THDM can be reduced to multiple applications of
the standard CP symmetry.}

Using Eq.~(\ref{X-prime}),
we see that the basis transformation of Eq.~(\ref{basis-transf}),
changes the standard CP symmetry of Eq.~(\ref{StandardCP})
into the GCP symmetry of Eq.~(\ref{GCP}),
with
\begin{equation}
X=U U^\top.
\end{equation}
In particular,
an orthogonal basis transformation does not affect the
form of the standard CP transformation.
Since we wish to generate $X \neq 1$,
we will need complex matrices $U$.

Now we wish to consider the following situation.
We have a basis (call it the original basis) and
impose the standard CP symmetry CP1 on that original basis.
Next we consider the same model in a different basis
(call it $M$) and impose the standard CP symmetry on that basis $M$.
In general, this procedure of imposing
the standard CP symmetry in the original basis \textit{and also}
in the rotated basis $M$ leads to two independent impositions.
The first imposition makes all parameters real in
the original basis.
One way to combine the second imposition with the first
is to consider the basis transformation $U_M$ taking us
from basis $M$ into the original basis.
As we have seen,
the standard CP symmetry in basis $M$ turns,
when written in the original basis,
into a symmetry under
\begin{eqnarray}
\Phi^{\textrm{CP}}_a
&=& (X_M)_{a \alpha} \Phi_\alpha^\ast,
\nonumber\\
\Phi^{\dagger \textrm{CP}}_a
&=&
(X_M)^\ast_{a \alpha} (\Phi_\alpha^\dagger)^\ast,
\label{CGP-M}
\end{eqnarray}
with $X_M=U_M U_M^\top$.
Next we consider several such possibilities.

We start with
\begin{equation}
U_{A} =
\left(
\begin{array}{cc}
\phantom{-}c_{\pi/4} & \quad -i s_{\pi/4} \\
- i s_{\pi/4} & \quad \phantom{-}c_{\pi/4}
\end{array}
\right),
\hspace{3ex}
X_{A} =
\left(
\begin{array}{cc}
\phantom{-}0 & \quad -i  \\
- i & \quad \phantom{-}0
\end{array}
\right).
\end{equation}
Here and henceforth $c$ ($s$) with a subindex indicates the
cosine (sine) of the angle given in the subindex.
We denote by CP1$_A$ the imposition of the CP symmetry
in Eq.~(\ref{CGP-M}) with $X_M=X_A$
(which coincides with the imposition of the standard CP
symmetry in the basis $M=A$).

Next we consider
\begin{equation}
U_{B} =
\left(
\begin{array}{cc}
e^{-i \pi/4} & \quad 0 \\
0 & \quad e^{i \pi/4}
\end{array}
\right),
\hspace{3ex}
X_{B} =
\left(
\begin{array}{cc}
-i & \quad 0  \\
\phantom{-}0 & \quad i
\end{array}
\right).
\end{equation}
We denote by CP1$_B$ the imposition of the CP symmetry
in Eq.~(\ref{CGP-M}) with $X_M=X_B$
(which coincides with the imposition of the standard CP
symmetry in the basis $M=B$).

A third possible choice is
\begin{equation}
U_{C} =
\left(
\begin{array}{cc}
e^{i \delta/2} & \quad 0 \\
0 & \quad e^{-i \delta/2}
\end{array}
\right),
\hspace{3ex}
X_{C} =
\left(
\begin{array}{cc}
e^{i \delta} & \quad 0 \\
0 & \quad e^{-i \delta}
\end{array}
\right),
\end{equation}
where $\delta \neq n \pi/2$ with $n$ integer.
We denote by CP1$_C$ the imposition of the CP symmetry
in Eq.~(\ref{CGP-M}) with $X_M=X_C$
(which coincides with the imposition of the standard CP
symmetry in the basis $M=C$).

Finally, we consider
\begin{equation}
U_{D} =
\left(
\begin{array}{cc}
\phantom{i}c_{\delta/2} & \quad i s_{\delta/2} \\
i s_{\delta/2} & \phantom{i}\quad c_{\delta/2}
\end{array}
\right),
\hspace{3ex}
X_{D} =
\left(
\begin{array}{cc}
\phantom{i}c_\delta & \quad i s_\delta  \\
i s_\delta & \quad \phantom{i}c_\delta
\end{array}
\right),
\end{equation}
where $\delta \neq n \pi/2$ with $n$ integer.
We denote by CP1$_D$ the imposition of the CP symmetry
in Eq.~(\ref{CGP-M}) with $X_M=X_D$
(which coincides with the imposition of the standard CP
symmetry in the basis $M=D$).

The impact of the first three symmetries on the coefficients of the
Higgs potential are summarized in
Table~\ref{master3}.
%
\begin{table}[ht!]
\caption{Impact of the CP1$_M$ symmetries on the coefficients
of the Higgs potential.
The notation ``imag'' means that the
corresponding entry is purely imaginary.
CP1 in the original basis has been included for reference.}
\begin{ruledtabular}
\begin{tabular}{ccccccccccc}
symmetry & $m_{11}^2$ & $m_{22}^2$ & $m_{12}^2$ &
$\lambda_1$ & $\lambda_2$ & $\lambda_3$ & $\lambda_4$ &
$\lambda_5$ & $\lambda_6$ & $\lambda_7$ \\
\hline
CP1 &  &  & real &
 &  & &  &
real & real & real\\
\hline
CP1$_A$ &   & $m_{11}^2$  &  &
   & $\lambda_1$ &  &  &
   &  & $\lambda_6$ \\
CP1$_B$ &  &  & imag &
 &  & &  &
real & imag & imag \\
CP1$_C$ &  &  & $|m_{12}^2| e^{i \delta}$ &
 & & & & $|\lambda_5| e^{2 i \delta}$  &
$|\lambda_6| e^{i \delta}$ & $|\lambda_7| e^{i \delta}$
\\
\end{tabular}
\end{ruledtabular}
\label{master3}
\end{table}
%

\noindent
Imposing CP1$_D$ on the Higgs potential leads to the more complicated
set of equations:
\begin{eqnarray}
2 \textrm{Im} \left( m_{12}^2 \right)\, c_\delta
+ (m_{22}^2 - m_{11}^2)\, s_\delta
&=& 0,
\nonumber\\
2 \textrm{Im} \left( \lambda_6 - \lambda_ 7 \right)\, c_{2 \delta}
+ \lambda_{12345}\, s_{2 \delta}
&=& 0,
\nonumber\\
2 \textrm{Im} \left( \lambda_6 + \lambda_ 7 \right)\, c_\delta
+ \left( \lambda_1 - \lambda_2 \right)\, s_\delta
&=& 0,
\nonumber\\
\textrm{Im} \lambda_5\, c_\delta
+ \textrm{Re} \left( \lambda_6 - \lambda_ 7 \right)\, s_\delta
&=& 0,
\end{eqnarray}
where
\begin{equation}
\lambda_{12345} = \tfrac{1}{2} \left( \lambda_1 + \lambda_ 2 \right)
- \lambda_3 - \lambda_4 + \textrm{Re} \lambda_5.
\end{equation}

Combining these results with those in Table~\ref{master1},
we have shown that
\begin{eqnarray}
\textrm{CP1} \oplus \textrm{CP1}_B
&=& Z_2\ \  \textrm{in some specific basis},
\nonumber\\
\textrm{CP1} \oplus \textrm{CP1}_C
&=& U(1),
\nonumber\\
\textrm{CP1} \oplus \textrm{CP1}_A \oplus \textrm{CP1}_B
&=& \textrm{CP2}\ \  \textrm{in some specific basis},
\nonumber\\
\textrm{CP1} \oplus \textrm{CP1}_A \oplus \textrm{CP1}_C
&=& \textrm{CP3}\ \  \textrm{in some specific basis},
\nonumber\\
\textrm{CP1} \oplus \textrm{CP1}_C \oplus \textrm{CP1}_D
&=& SO(3).
\label{incredible}
\end{eqnarray}
Let us comment on the ``specific basis choices'' needed.
Imposing $\textrm{CP1} \oplus \textrm{CP1}_B$ leads to
$ m_{12}^2=\lambda_6=\lambda_7=0$ and $\textrm{Im} \lambda_5=0$,
while imposing $Z_2$ leads to $ m_{12}^2=\lambda_6=\lambda_7=0$
with no restriction on $\lambda_5$.
However, when $Z_2$ holds one may rephase $\Phi_2$
by the exponential of $-i \arg(\lambda_5)/2$,
thus making $\lambda_5$ real.
In this basis,
the restrictions of $Z_2$ coincide with the restrictions
of $\textrm{CP1} \oplus \textrm{CP1}_B$.
Similarly,
imposing $\textrm{CP1} \oplus \textrm{CP1}_A \oplus \textrm{CP1}_C$
leads to $m_{12}^2=\lambda_5=\lambda_6=\lambda_7=0$,
$m_{22}^2=m_{11}^2$ and $\lambda_2=\lambda_1$.
We see from Table~\ref{master1} that CP3 has these features,
except that $\lambda_5$ need not vanish; it is real and
$\textrm{Re} \lambda_5 = \lambda_1-\lambda_3-\lambda_4$.
Starting from the CP3 conditions and
using the transformation rules in Eqs.~(A13)-(A23) of Davidson and
Haber \cite{DavHab},
we find that a basis choice is possible such that
$\textrm{Re} \lambda_5=0$.\footnote{Notice that, in the new basis,
$\lambda_1$ differs in general from $\lambda_3 + \lambda_4$;
otherwise the larger $SO(3)$ Higgs Family symmetry would hold.}
Perhaps it is easier to prove the equality
\begin{equation}
\textrm{CP1} \oplus \textrm{CP1}_B \oplus \textrm{CP1}_D
= \textrm{CP3}\ \  \textrm{in some specific basis}.
\end{equation}
In this case,
the only difference between the impositions from the
two sides of the equality come from the sign of $\textrm{Re} \lambda_5$,
which is trivial to flip through the basis change
$\Phi_2 \rightarrow - \Phi_2$.
Finally,
imposing $\textrm{CP1} \oplus \textrm{CP1}_A \oplus \textrm{CP1}_B$
we obtain $m_{12}^2=\textrm{Im} \lambda_5=\lambda_6=\lambda_7=0$,
$m_{22}^2=m_{11}^2$ and $\lambda_2=\lambda_1$.
This does not coincide with the conditions of CP2 which
lead to the ERPS of Eq.~(\ref{ERPS}).
Fortunately,
and as we mentioned before,
Davidson and Haber \cite{DavHab} proved that
one may make a further basis transformation
such that Eq.~(\ref{ERPS2}) holds,
thus coinciding with the conditions imposed by
$\textrm{CP1} \oplus \textrm{CP1}_A \oplus \textrm{CP1}_B$.

Notice that our description of CP2 in terms of several
CP1 symmetries is in agreement with the results found by the
authors of Ref.~\cite{mani}.
These authors also showed a very interesting
result, concerning  spontaneous symmetry breaking in 2HDM models
possessing a CP2 symmetry.
Namely, they prove (their Theorem 4)
that electroweak symmetry breaking will {\em necessarily}
spontaneously break CP2.
However, they also show that the vacuum
will respect at least one of the CP1 symmetries which compose
CP2.
Which is to say, in a model which has a CP2 symmetry,
spontaneous symmetry breaking necessarily respect the CP1
symmetry.

In summary,
we have proved that all possible symmetries on the scalar
sector of the THDM,
including Higgs Family symmetries,
can be reduced to multiple applications of
the standard CP symmetry.

\section{\label{sec:conclusions}Conclusions}

We have studied the application of generalized CP symmetries
to the THDM,
and found that there are only two independent classes
(CP2 and CP3),
in addition to the standard CP symmetry (CP1).
These two classes lead to an exceptional region of parameter,
which exhibits either a $Z_2$ discrete symmetry or
a larger $U(1)$ Peccei-Quinn symmetry.
We have succeeded in
identifying a basis-independent invariant quantity that can
distinguish between the $Z_2$ and $U(1)$ symmetries.
In particular, such an invariant is required
in order to distinguish between CP2 and CP3,
and completes the description of all symmetries in the THDM
in terms of basis-invariant quantities.
Moreover, CP2 and CP3 can be obtained by combining
two Higgs Family symmetries and that this is not possible
for CP1.  

We have shown that all symmetries of the THDM previously identified
by Ivanov \cite{Ivanov1} can be achieved through simple symmetries.
with the exception of $SO(3)$.
However, the $SO(3)$ Higgs Family symmetry
can be achieved by imposing a $U(1)$ Peccei-Quinn  
symmetry and the CP3-symmetry in the same basis.
Finally, we have demonstrated that
all possible symmetries of the scalar
sector of the THDM can be reduced to multiple applications of
the standard CP symmetry.
Our complete description of the symmetries on the scalar fields
can be combined with symmetries in the quark and lepton sectors,
to aid in model building.


\begin{acknowledgments}
We would like to thank Igor Ivanov and Celso Nishi for their
helpful comments on the first version of this manuscript.
The work of P.M.F. is supported in part by the Portuguese
\textit{Funda\c{c}\~{a}o para a Ci\^{e}ncia e a Tecnologia} (FCT)
under contract PTDC/FIS/70156/2006. The work of H.E.H. is
supported in part by the U.S. Department of Energy, under grant
number DE-FG02-04ER41268. The work of J.P.S. is supported in
part by FCT under contract CFTP-Plurianual (U777).

H.E.H. is most grateful for the kind hospitality and support of the
Centro de F\'{\i}sica Te\'orica e Computacional at Universidade de
Lisboa
(sponsored by the Portuguese FCT and
Funda\c{c}\~{a}o Luso-Americana para o Desenvolvimento)
and the Centro de F\'{\i}sica Te\'orica de Part\'{\i}culas at
Instituto Superior T\'ecnico during his visit to Lisbon. This work
was initiated during a conference in honor of Prof. Augusto Barroso,
to whom we dedicate this article.
\end{acknowledgments}

\end{document}